\documentclass[iop]{emulateapj}
\usepackage{natbib}
\shorttitle{Saturated torque formula for planetary migration}
\shortauthors{F.\ S.\ Masset \& J. Casoli}
\begin{document}
\title{Saturated torque formula for planetary migration in viscous disks with
thermal diffusion: recipe for protoplanet population synthesis}
\author{F.\ S.\ Masset\altaffilmark{1,2}} 
\affil{Laboratoire AIM,
 CEA/DSM - CNRS - Universit\'e Paris Diderot, 
 Irfu/Service d'Astrophysique, B\^at. 709,
 CEA/Saclay, 91191 Gif-sur-Yvette, France}
\altaffiltext{1}{On leave from Service d'Astrophysique, CE-Saclay,
  France. Now at Instituto de Ciencias F\'\i sicas, Universidad
  Nacional Aut\'onoma de M\'exico (UNAM), Apdo. Postal 48-3,
  62251-Cuernavaca, Morelos, M\'exico}
\email{masset@fis.unam.mx}
\and 
\author{J. Casoli}
\affil{Laboratoire AIM,
  CEA/DSM - CNRS - Universit\'e Paris Diderot, 
  Irfu/Service d'Astrophysique, B\^at. 709, 
  CEA/Saclay, 91191 Gif-sur-Yvette, France}
\email{jules.casoli@cea.fr}
\altaffiltext{2}{Send offprint requests to masset@fis.unam.mx}
\begin{abstract}
  We provide torque formulae for low mass planets undergoing type~I
  migration in gaseous disks. These torque formulae put special
  emphasis on the horseshoe drag, which is prone to saturation: the
  asymptotic value reached by the horseshoe drag depends on a
  balance between coorbital dynamics (which tends to cancel out or
  saturate the torque) and diffusive processes (which tend to restore
  the unperturbed disk profiles, thereby desaturating the torque). We
  entertain here the question of this asymptotic value, and we derive
  torque formulae which give the total torque as a function of
  the disk's viscosity and thermal diffusivity. The horseshoe drag
  features two components: one which scales with the vortensity
  gradient, and one which scales with the entropy gradient, and which
  constitutes the most promising candidate for halting inward type~I
  migration.  Our analysis, which is complemented by numerical
  simulations, recovers characteristics already noted by numericists,
  namely that the viscous timescale across the horseshoe region must
  be shorter than the libration time in order to avoid saturation,
  and that, provided this condition is satisfied, the entropy related
  part of the horseshoe drag remains large if the thermal timescale is
  shorter than the libration time. Side results include a study of the
  Lindblad torque as a function of thermal diffusivity, and a
  contribution to the corotation torque arising from vortensity
  viscously created at the contact discontinuities that appear at the
  horseshoe separatrices.  For the convenience of the reader mostly
  interested in the torque formulae, section~8 is self-contained.
\end{abstract}
\keywords{Planetary systems: formation --- planetary systems:
 protoplanetary disks --- Accretion, accretion disks --- Methods:
 numerical --- Hydrodynamics}
\section{Introduction\label{sec:intro}}
Ever since the discovery of the first extrasolar planets, many efforts
have been undertaken to account for the statistics of their orbital
properties, and to link these statistics to the properties of the
protoplanetary disk. In particular, the so-called planetary population
synthesis models correspond to a class of essentially one dimensional
models, in which the physics of the disk is treated so as to include,
in a simplified manner, as many relevant physical processes as
necessary, and in which the growth and migration of randomly sorted
planetary embryos is tracked until the disk disappears
\citep{2008Sci...321..814T,Mordasini:2009p398,Ida:2008p297,
  Ida:2008p396,Ida:2004p394,Ida:2004p393,Ida:2005p395,
  Alibert:2005p36,2007ApJ...656L..25T,2006ApJ...644.1214T}.  The
statistical properties inferred from these models can be confronted to
observational data, and thus allow to constrain the underlying disk
physical properties or the physics of the interaction of the embryos
with the disk. One drawback of these models, however, is that they
have to rely on tidal torque expressions, since the processes that
give rise to exchange of angular momentum between a planetary embryo
and the disk cannot be captured by a one dimensional analysis.

A problematic phase arises in these models, that corresponds to the
migration of low mass objects (the type~I migration phase). Standard
torque estimates, based on the linear expressions of \citet{w97} or
\citet{tanaka2002}, yield a rapid flush of the low mass cores to the
innermost parts of the protoplanetary disk. Several authors have
circumvented this problem by arbitrarily lowering the type~I migration
estimate by an {\em ad hoc} factor $f$ typically in the range $f\sim
0.01-0.1$, so as to allow a reasonable fraction of the initial cores
to remain as sizable distances of the star over the disk lifetime
\citep{Ida:2008p396,Mordasini:2009p399}.

The theory of the tidal interaction of low mass planetary objects with
the disk has recently changed, however, as \citet{2009arXiv0901.2265P}
have shown that one of the tidal torque components, the so-called
corotation torque, eventually becomes non-linear at all planetary
masses. This breakthrough renders obsolete the systematic use of the
linear estimate of \cite{tanaka2002} for low mass planets, and the
torque expression that should be used is the sum of the linear
Lindblad torque expression of \citet{tanaka2002}, and of the
non-linear corotation torque.  It corresponds to the torque exerted by
the coorbital region on to the planet. Only in the limit of low mass
and large diffusion, does the corotation torque approaches its linear
estimate. Far from this limit, the non-linear corotation amounts to
the horseshoe drag, which corresponds to the torque exerted by fluid
elements undergoing a horseshoe like libration in the coorbital
region, in the frame corotating with the planet.

The horseshoe drag bears a number of similarities with the linear
corotation torque: its scaling with all parameters (planet mass, disk
surface density, disk aspect ratio, etc.) is the same, but it turns
out to have in general a larger value. Another important difference is
its long-term behavior: the horseshoe dynamics tends to redistribute
angular momentum between the horseshoe region and the planet, but the
total angular momentum content of the system composed of the planet
and its coorbital region remains constant in an inviscid disk, if one
disregards what is carried away by the wake. As a consequence, the
horseshoe drag tends to zero on the long term, and the total angular
momentum ever exchanged between the disk and a planet initially
``switched on'' in the unperturbed disk is simply the difference of
angular momentum content of the horseshoe region between its final
state and its initial, unperturbed state. This process, known as
saturation, occurs on a relatively small time scale, typically of
$O(10^2)$ orbital periods for Earth-sized objects. It is in particular
much shorter than the disk lifetime and the migration time scale
itself, hence models that follow the migration of a core up to the
disk dispersal should consider the long term value of the horseshoe
drag (which we also call the asymptotic or saturated value), rather
than the value that it displays soon after the introduction of the
planet in an unperturbed disk (which we call the unsaturated
value). As stressed above, if there is no process that allows the
exchange of angular momentum between the horseshoe region and the rest
of the disk, the asymptotic horseshoe drag is simply zero. If the disk
has a non-vanishing viscosity, however, the viscous friction at the
boundaries of the horseshoe region (the separatrices) enables a steady
flux of angular momentum across the horseshoe region, and the
asymptotic value of the horseshoe drag may remain finite. It is the
purpose of this paper to provide the asymptotic value of the torque.

An estimate of the saturated horseshoe drag already exists in the
literature \citep{masset01}, but it applies exclusively to isothermal
(or barotropic) disks, that is to say to disks in which the pressure
is solely a function of the density. It has recently been shown that
the horseshoe dynamics is more complex in non-isothermal flows, which
obey an energy equation rather than a barotropic closure relation
\citep{pm06,bm08,pp08,mc09,2009arXiv0909.4552P}.

In any case, the key quantity that determines the value of the
horseshoe drag is the vortensity, or potential vorticity. It is the
value of the vertical component of the vorticity divided by the
surface density of the disk. Regardless of the equation of state, the
horseshoe drag always reduces to a simple integral involving the
vortensity distribution across the horseshoe region. In a
non-isothermal flow, baroclinic effects act as source or sink terms
for the vortensity, which in turn has a strong impact on the horseshoe
drag.  These effects have been found to be localized exclusively at
the separatrices of the horseshoe region \citep{mc09}, and they scale
with the entropy gradient in the disk. As entropy follows the
horseshoe streamlines and tends to be flattened over the horseshoe
region by phase mixing, the entropy gradient that persists over long
time scales across the horseshoe region depends on diffusive processes
that affect the evolution of entropy (such as thermal diffusion, and
in a more general manner all the processes that act as sink or source
terms in the energy equation). The asymptotic torque expression
therefore involves two main dissipative processes: viscosity, and
thermal diffusion, and it depends on two gradients: the gradient of
vortensity, and the gradient of entropy.

As a point of terminology, we mention that the component of the
horseshoe drag which scales with the vortensity gradient has been
called vortensity-related torque \citep{2009arXiv0909.4552P}, or bulk
term \citep{mc09}, whereas its additional non-barotropic component has
been called entropy-related torque \citep{2009arXiv0909.4552P}, edge
term \citep{mc09}, or adiabatic torque excess \citep{mc09,bm08}. We
will use indifferently these denominations, but we shall reserve the
last one for the unsaturated horseshoe drag, for which the dissipative
processes are unimportant and the flow can be considered as adiabatic.

This paper is organized as follows: section~\ref{sec:prerequisite}
presents the framework and introduces the notation and governing
equations, section~\ref{sec:simpl-model-flow} presents a reduced model
of the flow that retains its essential characteristics for the
horseshoe dynamics, section~\ref{sec:hors-drag-expr} presents a
derivation of the horseshoe drag expression that we shall use to get
the asymptotic estimates. This expression is strictly equivalent to
the expression of \citet{mc09}, in which the edge term is now
incorporated into the integral over the vortensity, which is singular
at the separatrix. In section~\ref{sec:numer-simul}, we present the
set of full hydrodynamical simulations that we have performed with the
FARGO code, and we compare their results with those of the reduced
model of section~\ref{sec:simpl-model-flow}. In
section~\ref{sec:saturation-functions}, we work out suitable
relationships for the steady state flow, and infer the corresponding
asymptotic horseshoe drag. Our analysis is exact in the limit of small
viscosity or small thermal diffusion, but its overall accuracy is
found to be satisfactory up to the values for which the torque remains
fully unsaturated. For the convenience of the reader mainly interested
in the results and their implementation,
section~\ref{sec:gener-torq-expr} is self-contained (the notation
required to follow that section is given in Tab.~\ref{tab:notation}).
It provides the torque value as a function of viscosity and thermal
diffusivity for low-mass planets subject to type~I
migration. Section~\ref{sec:discussion} provides additional
discussion, and we draw our conclusions in section~\ref{sec:summary}.

\section{Prerequisite}
\label{sec:prerequisite}
\subsection{Framework}
\label{sec:framework}
A fundamental premise of our analysis is that a turbulent
protoplanetary disk can be adequately modeled by a laminar,
non-magnetized disk, in which diffusion (of energy and angular
momentum) is described by phenomenological diffusion coefficients.
Turbulence, which likely results from the magnetorotational
instability \citep{bh91}, acts at exchanging angular momentum across
the horseshoe separatrices. We assume that it does so in a diffusive
manner on the radial scale of the horseshoe region. It is not clear
whether this is the case for the low or intermediate mass planets
considered here, for which the horseshoe zone width is at most of the
order of the disk thickness, that is to say of same order than the
largest scale of the turbulence. Since this is a totally unexplored
regime, which awaits dedicated three dimensional MHD calculations, we
shall restricts ourselves to this assumption so that it can be modeled
by resorting to the Navier Stokes equation with an effective kinematic
viscosity.  Little is known on the validity of this assumption in the
context of horseshoe drag saturation. Nevertheless, we mention the
encouraging results of \citet{2010ApJ...709..759B}, who have
considered an inviscid two-dimensional disk subject to stochastic
forcing, and found saturation properties of the horseshoe drag very
similar to those in a laminar disk.

In this picture the details of the properties of
turbulence should not matter. The scale at which the turbulent cascade
dissipates (resistive or viscous) may impact the saturated state of
the turbulence, i.e. the value of the effective viscosity
\citep[see][and refs. therein]{2010EAS....41..167F}. However,
provided the correct value of the effective viscosity is used, the
time averaged distribution of vortensity within the horseshoe region
should be insensitive to the statistical properties of the turbulence.

We note that all what matters to assess the asymptotic horseshoe drag
is how much angular momentum exchange subsists at large time between
the disk and the horseshoe region, considered here in a wide meaning
as everything enclosed within the separatrices, including the planet
\citep[e.g.][]{masset01}. For this reason, the horseshoe drag is
always linked to the flow properties at the separatrices. The internal
details of the coorbital flow, such as the existence of tadpole
streamlines or the topology of the flow in the vicinity of the planet,
do not impact the horseshoe drag. Stated differently, since the whole
set of librating material is trapped in the coorbital region and
migrates along with it \citep[e.g.][]{mp03}, the forces internal to
the horseshoe region are internal to the migrating system and
therefore do not affect its migration rate.

Turbulence also yields a stochastic torque that triggers a
random walk of the semi-major axis of the planet \citep{nelson05},
which comes in addition to the systematic drift due to the wake.
As a consequence the probability density of the planet's semi-major
axis obeys a diffusion-advection law \citep{jgm06}. When averaging over a
sufficient amount of time, the systematic drift overcomes the
stochastic effects. Although it is still unclear whether the drift so
averaged amounts to the drift obtained from a laminar analysis
\citep{nelson05}, we assume
here that this is the case, and that the long-term effects of
turbulence exclusively amount to regulating the degree of saturation of
the horseshoe region, from which we infer laminar torque values which
dictate the time averaged migration rate.

In a similar vein, a number of different processes, in addition to
turbulence itself, lead to a diffusion of temperature in the disk,
such as thermal diffusion or radiative diffusion, while cooling
through the disk's photosphere and heating by turbulence also
contribute to the energy budget.  For the sake of simplicity, we
shall assume that the time evolution of the fluctuations of
temperature with respect to the unperturbed state can be described by
a unique diffusion coefficient, to which we refer as the thermal
diffusivity, which is an effective diffusion coefficient much as the
effective viscosity. This is a rather standard practice
\citep[e.g.][]{pp08}. We shall see that all what matters to determine
the degree of saturation of the entropy related torque is the entropy
difference on both separatrices upstream of the U-turns, and that the
only parameter that determines this difference is the ratio of the
thermal timescale to the horseshoe libration timescale. Our
simplifying assumption therefore retains all the physics relevant to
the description of the saturation of the non-isothermal horseshoe
drag.

Finally, most of our study (analytical and numerical) is performed
assuming a two-dimensional disk, for reasons of tractability and
computational cost. In section~\ref{sec:transition-from-two} we will
try and generalize to three dimensions in the framework of stacked
two dimensional layers.

\subsection{Notation}
\label{sec:notation}
In addition to the notation defined below, the notation used in
section~\ref{sec:gener-torq-expr} is presented in
table~\ref{tab:notation}, for the convenience of the reader mainly
interested in the results presented in that section.  We consider a
planet of mass~$M_p$ orbiting a star of mass~$M_*$, on a fixed
circular orbit. This planet is embedded in a gaseous protoplanetary
disk, such that the planetary orbit is coplanar with the disk and
prograde. We use $P$ to denote the vertically integrated pressure, and
$s$ to denote the gas entropy, which reads:
\begin{equation}
 \label{eq:1}
 s = \frac{P}{\Sigma^\gamma},
\end{equation}
where $\Sigma$ and $\gamma$ are defined in Tab.~\ref{tab:notation}.

\begin{deluxetable*}{ccc}
\tablecaption{Main notation used in this work.\label{tab:notation}}
\tablehead{\colhead{Name} & \colhead{Notation} & \colhead{Comment}}
\startdata
Planet orbital frequency & $\Omega_p$&\\
Planet semi major axis & $a$&\\
Planet to star mass ratio & $q$ &$q=M_p/M_*$\\
Disk surface density &$\Sigma$ & $\Sigma=\Sigma_c(r/a)^{-\alpha}$\\
Vertically averaged temperature &$T$ & $T=T_c(r/a)^{-\beta}$\\
Flaring Index & $f$& $f=(1-\beta)/2$\\
 & & $h \propto r^f$\\
Disk aspect ratio & $h$ & $h=h_0(r/a)^f$\\
Vortensity gradient & ${\cal V}$ & ${\cal V} = 3/2-\alpha$\\ 
Adiabatic index & $\gamma$ &\\ 
Entropy gradient & ${\cal S}$ & ${\cal S} =
\frac{\gamma-1}{\gamma}\alpha+\frac{2f-1}{\gamma}$\\
& &The entropy scales as $r^{\gamma{\cal S}}$\\
Horseshoe half width & $x_s$ & \\
\enddata
\tablecomments{The entropy gradient $\xi$ defined by
  \citet{2009arXiv0909.4552P} is equivalent to $\gamma{\cal S}$.}
\end{deluxetable*}

We identify a location in the disk by its distance $r$ to the star and
its azimuth $\phi$ with respect to the planet. The gas has a radial
velocity $v_r$ and an azimuthal velocity $v_\phi$ in the frame
corotating with the planet (hence its angular frequency in an inertial
frame is $\Omega=v_\phi/r+\Omega_p$). The radius of corotation $r_c$
is the location in the unperturbed disk where the material has same
angular frequency as the planet: $\Omega(r_c)=\Omega_p$. We will make
use of the distance~$x$ to corotation: $x=r-r_c$. We also consider the
disk pressure scaleheight $H=c_s/\Omega=rh$.

We denote with a subscript $0$ the value of a variable in the
unperturbed disk, and with a subscript $1$ the perturbation of this
variable. Whenever we refer to the unperturbed value of a variable at
corotation, we use a $c$ subscript.

Using the notation of \cite{bm08}, we define the gradient of the
inverse of vortensity across corotation as:
\begin{equation}
  \label{eq:2}
  {\cal V}=\left.\frac{d\log\left(\Sigma_0/\omega_0\right)}{d\log r}\right|_c=\frac
  32+\frac {r_c}{\Sigma_c}
\left.\frac{d\Sigma_0}{dr}\right|_{r_c}=\frac 32-\alpha,
\end{equation}
where $\alpha$ is the index of the surface density power law (see
Tab.~\ref{tab:notation}), and where
$\omega=(1/r)\partial_r(r^2\Omega)$ is the vertical component of the
flow vorticity.  We will also make use of the first Oort's constant
$A=(1/2)rd\Omega/dr$, which quantifies the amount of shear in the
unperturbed flow, and of the second Oort's constant
$B=(1/2r)d(r^2\Omega)/dr$, which is half the vertical component of the
unperturbed flow's vorticity.  Instead of the vortensity $w$ defined
by $w=\omega/\Sigma$, we oftentimes consider its inverse $l=1/w$, that
we call the load for the sake of definiteness: $l=\Sigma/\omega$.  We define the gradient of
entropy across corotation as:
\begin{equation}
  \label{eq:3}
  {\cal S}=\left.\frac 1\gamma\frac{d\log s_0}{d\log r}\right|_c.
\end{equation}
We shall make use of the dimensionless variables
\begin{equation}
  \label{eq:4}
  L=(l-l_c)/l_c\mbox{~~~and~~~}S=(s-s_c)/s_c.
\end{equation}
We denote the slope of these variables in the unperturbed disk
respectively with $L_\infty'$ and $S_\infty'$, where the prime stands
for the derivative with respect to $x$. We have the following relationships:
\begin{eqnarray}
  \label{eq:5}
  L_\infty' &=& \frac{{\cal V}}{a}\\
 \label{eq:6}
  S_\infty' &=& \frac{\gamma{\cal S}}{a}.
\end{eqnarray}
The reason for the $\infty$ symbol in the expressions above will
appear in section~\ref{sec:bound-cond-reduc}. They represent the large
scale gradient in the unperturbed disk, and therefore the asymptotic
value of the slope of the load or entropy as one goes away from the
coorbital region.
\subsection{Governing equations}
\label{sec:governing-equations}
The governing equations of the flow are the equation of continuity,
the Navier-Stokes equations and the energy equation, together with the
closure relationship provided by the equation of state, which is that
of an ideal gas:
\begin{equation}
  \label{eq:7}
P={\cal R}\Sigma T/\mu,  
\end{equation}
where ${\cal R}$ is the ideal gas constant and $\mu$ the mean
molecular weight of the gas.  The equation of continuity reads, in the
frame corotating with the planet:
\begin{equation}
 \label{eq:8}
 \partial_t\Sigma+\frac 1r\partial_r(\Sigma rv_r)+\frac
 1r\partial_\phi(\Sigma v_\phi)=0.
\end{equation}
The Navier-Stokes equations read, respectively in the radial and
azimuthal directions:
\begin{equation}
\label{eq:9}
 \partial_tv_r+v_r\partial_rv_r+\frac{v_\phi}{r}\partial_\phi
v_r-r\Omega_p^2-2\Omega_pv_\phi-\frac{v_\phi^2}{r}
=-\frac{\partial_rP}{\Sigma}-\partial_r\Phi
+\frac{f_r}\Sigma,
\end{equation}
and:
\begin{equation}
\label{eq:10} 
D_tj=-\frac{\partial_\phi P}{\Sigma}-\partial_\phi\Phi+\frac{rf_\phi}{\Sigma},
\end{equation} 
where
$D_t\equiv \partial_t+v_r\partial_r+\frac{v_\phi}{r}\partial_\phi$ and
\begin{equation}
\label{eq:11}
j=r^2\Omega
\end{equation}
is the specific angular momentum. In Eqs.~(\ref{eq:9})
and~(\ref{eq:10}), $f_r$ and $f_\phi$ are respectively the radial and
azimuthal component of the viscous force per unitary surface. These
viscous forces are derived from the viscous stress tensor as follows:
\begin{eqnarray}
f_r &=& \frac{1}{r} \frac{\partial (r \tau_{rr})}{\partial r}  +
\frac{1}{r} \frac{\partial \tau_{r \phi}}{\partial \phi} -
\frac{\tau_{\phi \phi}}{r} \label{eq:12} \\
f_{\phi} &=& \frac{1}{r} \frac{\partial (r \tau_{\phi r})}{\partial r} +
\frac{1}{r} \frac{\partial \tau_{\phi \phi}}{\partial \phi} +
\frac{\tau_{r \phi}}{r}, \label{eq:13}
\end{eqnarray}
where the components of the viscous stress tensor are:
\begin{eqnarray}
\label{eq:14}
\tau_{rr} & = & 2 \eta D_{rr} - \frac{2}{3} \eta \nabla . {\bf v} \\ 
\label{eq:15}
\tau_{\phi \phi} & = & 2 \eta D_{\phi \phi} - \frac{2}{3} \eta \nabla .
{\bf v}
\\ 
\tau_{r \phi} & = & \tau_{\phi r} = 2 \eta D_{r \phi}, \label{eq:16}
\end{eqnarray}
where
\begin{eqnarray}
\label{eq:17} 
D_{rr} &=& \frac{\partial v_r}{\partial r}, D_{\phi \phi} = \frac{1}{r}
\frac{\partial v_{\phi}}{\partial \phi} + \frac{v_r}{r} \\
\label{eq:18} 
D_{r \phi} &=& \frac{1}{2} \left[ r \frac{\partial}{\partial r} \left(
\frac{v_{\phi}}{r} \right) + \frac{1}{r} \frac{\partial v_r}{\partial \phi}
\right],
\end{eqnarray}
and $\eta=\Sigma \nu$ is the vertically integrated dynamical viscosity
coefficient.  The internal energy density is $e=p/(\gamma-1)$, and the
energy equation reads:
\begin{equation}
 \label{eq:19}
 \Sigma D_t\left(\frac e\Sigma\right)=-p\vec\nabla.\vec v
 -\vec\nabla.\vec F,
\end{equation}
where the last term of the right hand side accounts for the energy
diffusion, and will be given explicitly in section~\ref{sec:time-evol-entr}.

\subsection{Time evolution of the load}
\label{sec:time-evolution-load}
The evolution of the load or vortensity that the material undergoes in
the coorbital region corresponds to a slow transformation which takes
place while the fluid elements describe essentially circular
trajectories, far from the planet, between successive horseshoe
U-turns.  We therefore neglect the azimuthal derivatives in the
equation governing the evolution of the load. Eq.~(\ref{eq:10}) can
then be recast as:
\begin{equation}
  \label{eq:20}
  \partial_tj+v_r r\omega = \frac{1}{r\Sigma}\partial_r(r^2\tau_{\phi r}),
\end{equation}
where we have used (\ref{eq:11}), (\ref{eq:13}) and~(\ref{eq:16}).
Deriving with respect to $r$, using Eq.~(\ref{eq:8}) and~(\ref{eq:18}), we are left
with:
\begin{equation}
  \label{eq:21}
  D_tw
  =\frac{1}{r\Sigma}\partial_r\left[\frac{1}{r\Sigma}\partial_r(\Sigma\nu r^3\partial_r\Omega)\right].
\end{equation}
We assume that $\nu$ is a power law of radius, such that the inner
bracket of Eq.~(\ref{eq:21}) is flat in an unperturbed disk, which
amounts to assuming that there is no radial drift of material. We
mention that, in some of our numerical simulations, we have precisely
adopted such a prescription (see section~\ref{sec:runs-with-an}). We
can thus write, keeping only the highest order derivatives of the
perturbed quantities:
\begin{equation}
  \label{eq:22}
D_tw\approx \nu\partial^2_{r^2}w_1-2\nu
w_0\frac{\partial^2_{r^2}\Sigma_1}{\Sigma_0},
\end{equation}
where $\Omega_0$ and $\Sigma_0$ are respectively the angular velocity
and surface density of the unperturbed disk, whereas $\Omega_1$ and
$\Sigma_1$ are the perturbations of azimuthal velocity and surface
density.
In Eq.~(\ref{eq:22}) 
we have made use of the first order expansion:
\begin{equation}
  \label{eq:23}
  w_1 \approx r\frac{\partial_r\Omega_1}{\Sigma_0}-\frac{w_0}{\Sigma_0}\Sigma_1,
\end{equation}
and of the relationship, specific to Keplerian disks:
\begin{equation}
  \label{eq:24}
\partial_r\Omega_0=-3\frac{w_0\Sigma_0}{r}.  
\end{equation}
Recasting Eq.~(\ref{eq:22}) in terms of the load, we obtain:
\begin{equation}
  \label{eq:25}
  D_tl\approx \nu\partial^2_{r^2} l+2\nu
l_0\frac{\partial^2_{r^2}\Sigma}{\Sigma_0},
\end{equation}
where we have used the relationships
$\partial^2_{r^2}l_1\approx \partial^2_{r^2}l$ and
$\partial^2_{r^2}\Sigma_1\approx \partial^2_{r^2}\Sigma$. The typical
value of the perturbed load in the horseshoe region is indeed
$O(l_cx_s/a$), whereas it varies radially over a length scale at most
equal to $x_s$. The same applies to the surface density, except in a
barotropic disk: we note that if the radial scale of the disturbances
in the disk is small with respect to the pressure scale length
(i.e. with respect to the disk thickness), then, in a barotropic disk,
the excitation of evanescent pressure waves renders the relative
perturbation of surface density much smaller than the relative
perturbation of vortensity or load \citep{cm09}, so that the
perturbation of vortensity can be put solely on the account of the
perturbation of vorticity, and Eq.~(\ref{eq:25}) can then be recast
as:
\begin{equation}
  \label{eq:26}
  D_tl \approx \nu\partial^2_{r^2} l.
\end{equation}
On the contrary if the pressure length scale is shorter than the
radial scale of the disturbances, then the acoustic spread of the
perturbations does not significantly alter the perturbations of
surface density, so that the relative perturbation of load is equal
to the relative perturbation of surface density. In that case,
Eq.~(\ref{eq:25}) yields:
\begin{equation}
  \label{eq:27}
    D_tl \approx 3\nu\partial^2_{r^2} l.
\end{equation}
In the case that we contemplate in this work, that of a low-mass
planet embedded in a disk, for which the half-width of the horseshoe
region is smaller than the disk thickness \citep{mak2006}, the regime
that prevails corresponds to Eq.~(\ref{eq:26}). Eq.~(\ref{eq:27})
would be valid for very massive protoplanets, but it would be of
little interest, since those clear a gap in their coorbital region,
thereby shutting off the corotation torque.

We note that in a non-barotropic disk, in which there can be entropy
waves and therefore contact discontinuities, the last term of
Eq.~(\ref{eq:25}) may no longer be negligible. This is in particular
the case at the outer edge of the horseshoe region, downstream of the
U-turns \citep{mc09}. We shall therefore use Eq.~(\ref{eq:26}) for
barotropic disks, and Eq.~(\ref{eq:25}) for disks with an energy
equation.

\subsection{Time evolution of the entropy}
\label{sec:time-evol-entr}
Using Eq.~(\ref{eq:7}) we can transform Eq.~(\ref{eq:19}) into:
\begin{equation}
  \label{eq:28}
  \frac{{\cal R}\Sigma T}{(\gamma-1)\mu} D_t\log s = -\vec
  \nabla.\vec F,
\end{equation}
where $\vec F$ is the heat flux, given by:
\begin{equation}
  \label{eq:29}
  F = -k\nabla T,
\end{equation}
where $k$ is the thermal diffusivity. Assuming that it can be regarded
as constant over the length scales over which $T$ varies,
Eq.~(\ref{eq:28}) can be recast as:
\begin{equation}
  \label{eq:30}
  \frac{{\cal R}\Sigma T}{(\gamma-1)\mu} D_t\log s = k\Delta T.
\end{equation}
We now note that, for the case we are interested in, that of a
horseshoe region bound by contact discontinuities, the variations of
temperature occur on a much smaller scale than those of pressure,
which occur on the disk scale height. We therefore make the
following simplifying assumption, using Eq.~(\ref{eq:1}):
\begin{equation}
  \label{eq:31}
  \frac{\Delta T}{T} = \frac{1}{\gamma}\frac{\Delta s}{s}.
\end{equation}
Using Eqs.~(\ref{eq:30}) and~(\ref{eq:31}), we are led to:
\begin{equation}
  \label{eq:32}
  D_ts = \kappa\Delta s,
\end{equation}
where
\begin{equation}
  \label{eq:33}
  \kappa = \frac{\gamma-1}{\gamma} \frac{k\mu}{{\cal R}\Sigma}.
\end{equation}
The simplifying assumption underlying Eq.~(\ref{eq:31})
has several important
properties:
\begin{itemize}
\item Our prescription for thermal diffusion has no impact on the
  acoustic waves, since the diffusion term scales with the
  perturbations of entropy, which vanish in acoustic waves. It has
  therefore no impact on the differential Lindblad torque, and allows
  to disentangle in a clean manner the variations of the corotation
  torque due to variations of thermal diffusion.
\item Eq.~(\ref{eq:32}) shows that the evolution of entropy does not
  depend on other variables, if we regard $\Sigma$ as a constant, and
  if we consider that the streamlines do not depend on the
  perturbation that builds up in the coorbital region. Finding a
  steady state solution for the coorbital flow can therefore be done
  in two subsequent steps: (i) one first seeks a steady state solution
  for the entropy, which does not require the knowledge of any other
  variable, (ii) one then seeks a steady state solution of the
  vortensity, as the latter depends on the entropy field, in a manner
  that shall be specified in the next section.
\end{itemize}
We comment that we have undertaken a side study of the variation of
the Lindblad torque as the thermal diffusivity varies (see section
\ref{sec:diff-depend-diff}) using the alternate form of the diffusion
term (which then scales with the Laplacian of the temperature, rather
than the entropy). The transition of the Lindblad torque from the
adiabatic limit to the isothermal limit (the respective torque values
differ by a factor $\gamma$) occurs for values of the thermal
diffusion much larger than those relevant for the saturation of the
entropy related corotation torque.

\section{A simplified model of the flow}
\label{sec:simpl-model-flow}
\subsection{Simplifying assumptions}
\label{sec:simpl-assumpt}
\begin{figure}
  \centering
  \plotone{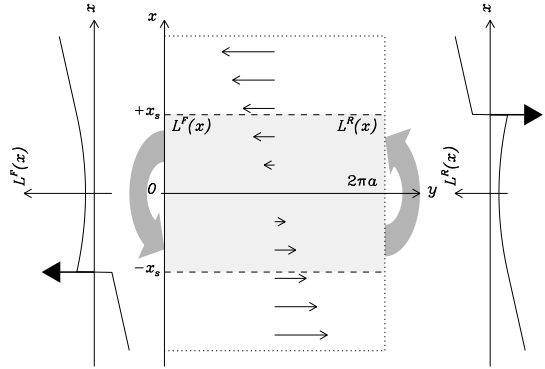}
  \caption{Sketch of the flow and of the load profile in front and
    behind the planet, within our simplified framework.  The graphs on
    the left and right side present a sketch of the front and rear
    load profiles, respectively.  The thick
    arrows in these graphs represent the singular load on the
    downstream separatrix. The light shaded area enclosed within the
    horizontal dashed lines shows the horseshoe region. The functions
    $L^F(x)$ and $L^R(x)$ are even for $x\in[-x_s,x_s]$.  The central
    set of arrows depicts the Keplerian flow, for which $v_y=2Ax$.}
  \label{sketch1}
\end{figure}

Owing to the complexity of the coorbital flow of an embedded planet,
we have to make a number of simplifying assumptions which render
tractable the task of finding the expression of the steady state
horseshoe drag. We anticipate that these assumptions do not impact
significantly the horseshoe drag expression. We will check this in
section~\ref{sec:comp-barotr-case}.
\begin{itemize} 
\item We separate the time scales of the horseshoe U-turns and that of
  the drift between two successive U-turns (i.e. half the libration
  time scale). The former is at least one order of magnitude shorter
  than the latter \citep{bm08}. This allows us to consider that the
  horseshoe U-turns are performed instantaneously.
\item Similarly, we consider that the azimuthal interval over which
  the U-turns take place is small compared to $2\pi$. This amounts to
  considering that the U-turns are squeezed against the axis joining
  the star and the planet.
\item We assume that the vortensity (or load) is conserved during the
  U-turns, except for the stagnation streamline (separatrix) in the
  non barotropic case, at which a singular amount of vortensity is
  created \citep{mc09}. Note that some vortensity is created during the
  U-turns in a non barotropic flow \citep{mc09} for all streamlines,
  but this has no impact on the horseshoe drag, hence we disregard it
  in this analysis. Also, the vortensity is ill-defined at the
  downstream separatrix, where there is both a contact discontinuity
  and a vorticity sheet. Nevertheless, the contact discontinuity has a
  surface density jump which is first order in $x_s/a$ \citep{bm08},
  hence to lowest order in $x_s/a$ the singularity of vortensity can
  be defined as $\Delta v_\phi\delta (x-x_s)/\Sigma_0$, where $\Delta
  v_\phi$ is the jump of azimuthal velocity across the downstream
  separatrix.
\item We assume that the streamlines can be regarded as circular
  between two U-turns.  This amounts to neglecting the radial velocity
  in the coorbital region, and to disregarding the behavior induced by
  the indirect term of the potential.
\item We assume that the U-turns are symmetric, i.e. that a streamline
  which has a radius $r$ prior to a U-turn is mapped to a streamline
  with radius $2r_c-r$, its distance to corotation remaining therefore
  $|r-r_c|$.
\end{itemize}
A consequence of this set of assumptions is that the path followed by
a horseshoe fluid element in the $(y, x)$ plane is rectangular (where
$y=r\phi$), and that the load and entropy are even functions of $x$
over the interval $[-x_s,x_s]$ in $\phi=0$ (front U-turns) and in
$\phi=2\pi$ (rear U-turns), except for the possible creation of a
singular amount of load in $x=\pm x_s$. Hereafter we will write with a
superscript $F$ ($R$) the quantities in $y=0^+$ ($y=2\pi a^-$) so as
to remind that they are evaluated in front (at the rear) of the
planet. A generic situation for the load is represented in
Fig.~\ref{sketch1}.

\subsection{Governing equations of the reduced model}
\label{sec:govern-equat-reduc}
In the framework considered in the previous section, we consider a
restricted model of the flow, in which we have only one scalar field
in the barotropic case (the load) or two scalar fields in the case
with an energy equation: the load and the entropy. In the barotropic
case, the variable $L$ obeys an equation derived from
Eq.~(\ref{eq:26}), which reads in the framework of the reduced model:
\begin{equation}
\label{eq:34}
\partial_tL+2Ax\partial_yL=\nu L''.
\end{equation}  
In the case with an energy equation, the variable $L$ obeys an
equation similar to Eq.~(\ref{eq:25}). The radial scale for the
variation of pressure is much larger than the scale of variation of
entropy or surface density, so we are led to:
\begin{equation}
\label{eq:35}
\partial_tL+2Ax\partial_yL=\nu L''-\frac{2\nu}{\gamma}S''.
\end{equation}  
In this case, we also need to consider the time evolution of the
entropy. Eq.~(\ref{eq:32}) yields:
\begin{equation}
 \label{eq:36}
 \partial_tS+2Ax\partial_yS=\kappa S''.
\end{equation}
Eqs.~(\ref{eq:34}) to~(\ref{eq:36}) amount to considering exclusively
the radial diffusion of these quantities, and to assuming that the
underlying flow is the unperturbed Keplerian flow, rectified on a
Cartesian slab.

\subsection{Boundary conditions of the reduced model}
\label{sec:bound-cond-reduc}
The boundary conditions of this reduced model are periodic in $y$
outside of the horseshoe region, with a period equal to $2\pi a$, that
is to say:
\begin{equation}
  \label{eq:37}
  L^F(x) = L^R(x)\mbox{~and~} S^F(x)=S^R(x)\mbox{~for any $x$ such that $|x|>x_s$},
\end{equation}
whereas within the horseshoe region the following relationships hold:
\begin{eqnarray}
\label{eq:38}
L^F(x)&=&L^F(-x)+L_0\delta(x+x_s) \mbox{~for any $x\in[-x_s,0]$}\\
 \label{eq:39}
L^R(x)&=&L^R(-x)-L_0\delta(x-x_s) \mbox{~for any $x\in[0,x_s]$}
\end{eqnarray}  
and
\begin{eqnarray}
\label{eq:40}
S^F(x)&=&S^F(-x) \mbox{~for any $x\in[-x_s,x_s]$}\\
 \label{eq:41}
S^R(x)&=&S^R(-x)\mbox{~for any $x\in[-x_s,x_s]$}
\end{eqnarray} 
Eqs.~(\ref{eq:38}) and~(\ref{eq:39}) are meant to account for the
horseshoe U-turns, in which the load is conserved, except at the
downstream separatrices where a singular amount $\pm \Lambda_0=L_0l_c$
is created, as has been shown by \citet{mc09}, and as we shall discuss
more in detail in section~\ref{sec:hors-drag-expr}.  The entropy obeys
similar equations, except that it is conserved during the U-turns,
even at the separatrix.

In the $x$ direction, boundary conditions are set for $|x|\rightarrow
\infty$ such that the slopes of the fields $L$ and $S$ are that of the
unperturbed flow:
\begin{equation}
 \label{eq:42}
 \lim_{x\rightarrow \pm\infty}\partial_xL(x,y) = L_\infty' \mbox{~~~for any $y\in[0,2\pi a]$}
\end{equation} 
and
\begin{equation}
 \label{eq:43}
 \lim_{x\rightarrow \pm\infty}\partial_x S(x,y) = S_\infty' \mbox{~~~for any $y\in[0,2\pi a]$}
\end{equation}

\subsection{Symmetry of the reduced model}
\label{sec:symm-reduc-model}
The reduced model exhibits the following symmetry:
\begin{equation}
 \label{eq:44}
 L(x,y) = -L(-x,2\pi a-y).
\end{equation}
while a similar property holds for the entropy field. The system is
set with this symmetry at $t=0$, since $L(x,y)\propto x$ in the
unperturbed flow, and the governing equation~(\ref{eq:34}) as well as
the boundary conditions specified by Eqs.~(\ref{eq:37}) to
(\ref{eq:39}) ensure that Eq.~(\ref{eq:44}) holds at any time. As a
consequence we have:
\begin{equation}
\label{eq:45}
L^R(-x)=-L^F(x)\mbox{~~~for any $x$}
\end{equation} 
and
\begin{equation}
\label{eq:46}
S^R(-x)=-S^F(x)\mbox{~~~for any $x$}
\end{equation}

\subsection{General considerations on the steady state of the reduced model}
\label{sec:gener-cons-steady}
The scalar field $S$ is not subject to any feed back from $L$: it
evolves by itself. On the contrary, the singular vortensity $\pm
\Lambda_0$ created at the downstream separatrices is linked to a
difference of the value of the entropy at the upstream front and rear
separatrices \citep{mc09}. We have, using Eq.~(62) of \citet{mc09}:
\begin{equation}
\label{eq:47}
\Lambda_0=-\frac{\Sigma_0\Delta v}{(2B)^2}=\frac{\Gamma_1}{16a|A|x_s^2B^2},
\end{equation}
where $\Gamma_1$, the adiabatic torque excess, has been worked out by
\citet{mc09} [see e.g.  their Eq.~(96)], and where $\Delta v$ is the
azimuthal velocity jump across the downstream separatrix.  In an
unsaturated situation, $\Gamma_1$ (and therefore $\Lambda_0$) depend
on the unperturbed entropy gradient ${\cal S}$ in the disk, which
imposes the entropy difference at the upstream separatrices. In a
saturated situation, the latter must be substituted by the effective
entropy gradient, which depends on the actual value of the entropy
difference at the upstream separatrices (rather than the unperturbed
one):
\begin{equation}
  \label{eq:48}
  {\cal
    S}'=\frac{1}{\gamma}\frac{a}{s_c}\frac{s^F(x_s)-s^R(-x_s)}{2x_s}=
\frac{1}{\gamma}\frac{a}{x_s}S^F(x_s),
\end{equation}
where we have used the symmetry property of Eq.~(\ref{eq:46}).  The
steady state situation is therefore determined as follows:
\begin{itemize}
\item A steady state solution is sought for the entropy field $S$.
\item Using Eq.~(\ref{eq:48}), this solution is used to evaluate
  $\Gamma_1$ and therefore $\Lambda_0$, the production of vortensity
  at the downstream separatrix.
\item Once $\Lambda_0$ is known, one can determine a steady state
  solution for the load field $L$.
\item The latter can then be used to evaluate the horseshoe drag as
  explained in section~\ref{sec:hors-drag-expr}.
\end{itemize}

\subsection{Implementation of the reduced model}
\label{sec:impl-reduc-model}
The reduced model presented above can be easily implemented as a
simple grid-based code. It allows to get values of the asymptotic
horseshoe drag at low computational cost. Namely, we solve the
evolution equations~(\ref{eq:34}) and (\ref{eq:36}) on a mesh
subdivided in $N_x\times N_y$ zones. The advection of the field
(either load or entropy) can be ensured by any upstream TVD scheme (in
our case we use a monotonized centered slope limiter). 
The timestep is limited by the Courant criterion, which reads here:
\begin{equation}
  \label{eq:49}
  \Delta t = C_0\min[(\Delta t_1^{-2}+\Delta t_2^{-2}+\Delta t_3^{-2})^{-1/2}],
\end{equation}
where $\Delta t_1 = \Delta x^2/(2\nu)$, $\Delta t_2 = \Delta
x^2/(2\kappa)$, $\Delta t_3 = \Delta y/|2Ax|$, $\Delta x$ and $\Delta
y$ being the mesh resolution respectively in $x$ and $y$ ($\Delta x
\ll \Delta y$). In Eq.~(\ref{eq:49}) we chose the Courant number
$C_0=0.9$, and the $\min$ operator is meant to be taken for all the
zones. The quadratic sum that features in Eq.~(\ref{eq:49}) is similar to
that in Eq.~(74) of \citet{zeus}.

The diffusion either of load or entropy is implemented by a second
order explicit operator, in the $x$ direction only. The
corresponding substep therefore reads, for the load:
\begin{equation}
  \label{eq:50}
  L^b_{ij} = L^a_{ij}+\nu\Delta t\frac{L^a_{i+1j}+L^a_{i-1j}-2L^a_{ij}}{\Delta x^2},
\end{equation}
where $i$ and $j$ are the zone indexes respectively in $x$ and $y$,
and where the $a$ and $b$ superscripts denote the load respectively
before and after the diffusion substep.

\begin{figure}
 \centering
 \plotone{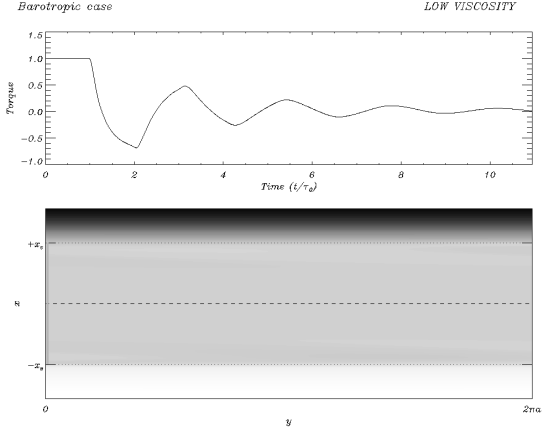}
 \caption{Still frame of the animation, that shows the load
   distribution at the end of the calculation (bottom) and the
   horseshoe drag as a function of time, in arbitrary units (top). The
   viscosity is low in this case, so that the load distribution
   eventually becomes flat. The animations shows two additional cases:
   a high viscosity case, in which the torque levels off to a
   non-vanishing value, and a case with an energy equation (i.e. the
   appearance of singular load sheets at the downstream separatrices)
   in which the torque oscillates and eventually decays to a rather
   low value.}
\label{fig:anim}
\end{figure}

The mesh is
designed so that the corotation and both separatrices fall on the
interface between adjacent rows of zones. The horseshoe region
therefore covers an integer and even number of rows.  One column of
ghost zones is used on each side in $y$. For the zones of the inner or
outer disk, standard periodic boundary conditions are applied in order
to set the values in the ghost zone. For the rows which are within the
horseshoe region, the boundary conditions differ, so as to mimic the
U-turns.  An animation of the time dependent load distribution is
available in a separate file. Fig.~\ref{fig:anim} shows a still frame
extracted from this animation, which also displays the corresponding
horseshoe drag value. The manner in which the latter is estimated is
detailed in the next section.

\section{Horseshoe drag expression}
\label{sec:hors-drag-expr}
We use the horseshoe drag expression given by \citet{mc09}, which we
transform so that the entropy related torque (corresponding to the
last term of this equation) feature the load produced during the
U-turns.  The horseshoe drag expression given at Eq.~(37) by
\citet{mc09} can be recast as:
\begin{equation}
  \label{eq:51}
  \Gamma=\Gamma^F+\Gamma^R,
\end{equation}
where the superscripts $F$ and $R$ respectively stand for the front
and rear contribution to the horseshoe drag, which read respectively:
\begin{eqnarray}
  \label{eq:52}
  \Gamma^F &=& \int_{G_s}^{G_c}l_+\Delta j_0dG-\frac{\Sigma_0}{4B}\Delta
  j_0^s\Delta G\\
  \Gamma^R &=& -\int_{G_s}^{G_c}l_-\Delta
  j_0dG-\frac{\Sigma_0}{4B}\Delta j_0^s\Delta G, \label{eq:53}
\end{eqnarray}
where $G$ is defined at Eq.~(16) of \citet{mc09}, and where $\Delta
j_0=4aBx$ is the jump of specific angular momentum of fluid elements
that execute a U-turn from $r=a+x$ to $r=a-x$, while $\Delta j^s_0$
refers to the same quantity for $x=x_s$. Note that we have split the
torque excess (trailing term of Eq.~(37) in MC09) into a front part
and a rear part, which are equal for symmetry reasons.  Using
Eqs.~(19), (37) and (62) of \citet{mc09}, one can recast the front
contribution to the horseshoe drag as an integral over $0<x<x_s$,
corresponding to the streamlines which undergo a horseshoe U-turn
towards smaller radii in front of the planet:
\begin{equation}
\label{eq:54}
\Gamma^F=\int_{0}^{x_s}4|A|Bl^F(x)\Delta j_0. xdx+2|A|B\Lambda_0\Delta j^s_0x_s,
\end{equation}
By convention, Eq.~(\ref{eq:54}) and all the torque expressions that
we shall write correspond to the torque exerted by the disk onto the
planet. From the definition of $\Delta j_0$ it follows that it is a
positive quantity on the interval considered, and that $\Gamma^F$ is
also positive, as expected since it stems from material in front of
the planet. In Eq.~(\ref{eq:54}), the trailing term comes from the
torque excess expression given by Eq.~(37) of \citet{mc09}, that we
have transformed using Eq.~(\ref{eq:47}).  We recast Eq.~(\ref{eq:54})
as:
\begin{equation}
  \label{eq:55}
\Gamma^F=16|A|B^2a\int_{0}^{x_s}l^F(x)x^2dx+8|A|B^2a\Lambda_0x_s^2,
\end{equation}
We transform Eq.~(\ref{eq:55}) using Eq.~(\ref{eq:38}), which allows
to embed the singular contribution within the integral:
\begin{equation}
 \label{eq:56}
\Gamma^F=8|A|B^2a\int_{-x_s}^{x_s}l^F(x)x^2dx.
\end{equation}
In this equation the integrand is singular in $x=-x_s$, that is to say
at the edge of the integration domain. It is therefore necessary to
specify explicitly how to deal with this singular
contribution. Comparison of Eqs.~(\ref{eq:55}) and~(\ref{eq:56}) shows
that one must take into account the whole singular distribution in the
integral (as if it was approximated by a narrow function whose support
was entirely contained within the interval of integration). Hereafter,
whenever we shall consider a horseshoe drag-like integral, with an
integrand singular on an edge (i.e. at a separatrix), we shall
implicitly consider that all the singularity must be taken into
account in the integral.

The rear contribution to the horseshoe drag can be worked out in a
similar manner and reads:
\begin{equation}
 \label{eq:57}
\Gamma^R=-8|A|B^2a\int_{-x_s}^{x_s}l^R(x)x^2dx,
\end{equation}
where, again, $l^R(x)$, the load at the rear of the planet, may
contain a singular contribution in $x=x_s$, arising from the
production of load during the U-turn.  The whole horseshoe drag
therefore reads:
\begin{equation}
  \label{eq:58}
  \Gamma=4|A|Ba\Sigma_0\int_{-x_s}^{x_s}[L^F(x)-L^R(x)]x^2dx,
\end{equation}
This expression can be further simplified, using Eq.~(\ref{eq:45}),
as:
\begin{equation}
  \label{eq:59}
  \Gamma=8|A|B\Sigma_0a\int_{-x_s}^{x_s}L^F(x)x^2dx,
\end{equation}
Simple particular cases can be recovered from Eq.~(\ref{eq:59}):
\begin{itemize}
\item In the unsaturated barotropic case with a vortensity gradient
  ${\cal V}=aL_\infty'$, we have $L^F(x)={\cal V}x/a$, from which we
  infer:
\begin{equation}
  \label{eq:60}
  \Gamma=4|A|B\Sigma_0{\cal V}x_s^4,
\end{equation}
or:
\begin{equation}
 \label{eq:61}
 \Gamma=\Gamma_0{\cal V},
\end{equation} 
which is the standard barotropic horseshoe drag expression
\citep{wlpi91,masset01}, where
\begin{equation}
 \label{eq:62}
 \Gamma_0=\frac 34\Sigma_0\Omega^2x_s^4,
\end{equation}
if we specialize to the case of a Keplerian disk.
\item In an inviscid barotropic disk in steady state in the corotating
  frame, the load is conserved along a streamline, hence $L^F(x) =
  L^R(x)$.  Since $L^R(x)=L^R(-x)=-L^F(x)$, we infer that $L^F(x)=0$
  for any $x\in[0,x_S]$. The horseshoe drag therefore cancels out. It
  is fully saturated, and the load is profile is flat across the
  horseshoe region.
\item In an adiabatic flow without vortensity gradient, the only
  contribution to the integrand stems from the singularity $\Lambda_0$
  at the downstream separatrix, which leads to:
\begin{equation}
  \label{eq:63}
  \Gamma=16|A|B^2ax_s^2\Lambda_0,
\end{equation}
which, using Eq.~(\ref{eq:47}), yields $\Gamma=\Gamma_1$.
\end{itemize}

\section{Numerical simulations}
\label{sec:numer-simul}
In addition to the calculations performed with the reduced model
presented in section~\ref{sec:simpl-model-flow}, we have performed
three series of full hydrodynamical simulations. We describe hereafter
the code and the different setups that we have adopted.

\subsection{Code and setup}
\label{sec:code-setup}
We have used the hydrocode FARGO\footnote{\tt http://fargo.in2p3.fr}
in its isothermal version \citep{fargo2000,fargo2000b} and in its
adiabatic version \citep{bm08,phdbaruteau}. The FARGO code is a
staggered mesh hydro-code on a polar grid, with upwind transport and a
harmonic, second order slope limiter \citep{1977JCoPh..23..276V}. It
solves the Navier-Stokes and continuity equations for a Keplerian disk
subject to the gravity of the central object and that of embedded
protoplanets, and the energy equation in the case of the adiabatic
version. It uses a change of rotating frame on each ring that enables
one to increase significantly the time step. The hydrodynamical solver
of FARGO resembles the widely known one of the ZEUS code \citep{zeus},
except for the handling of momenta advection. The Coriolis force is
treated so as to enforce angular momentum conservation
\citep{1998A&A...338L..37K}. The mesh is centered on the primary. It
is therefore non-inertial. The frame acceleration is incorporated in
the potential indirect term.

\subsubsection{Dissipation properties}
\label{sec:diss-prop}
\paragraph{Viscosity} 
\label{sec:viscosity}
The full viscous stress tensor in cylindrical coordinates of the
Navier-Stokes equations, given by Eqs.~(\ref{eq:12}) to (\ref{eq:18}),
is implemented in FARGO. As we shall see by comparing our results to
that of the reduced model, which features only the diffusion of load
due to the radial shear in Eq.~(\ref{eq:34}), the saturation
properties rely essentially on the radial shear, and the other
components of the viscous stress tensor appear to be unimportant.

\paragraph{Diffusivity}
\label{sec:diffusivity}
As discussed in section~\ref{sec:time-evol-entr}, we assume that, for
our purpose, the evolution of the entropy can be described by a
diffusion equation. Since the entropy is not one of the primitive
variables handled by the hydrodynamical solver, the thermal diffusion
substep updates the internal energy. This substep amounts to solving
the following partial differential equation:
\begin{equation}
  \label{eq:64}
  \partial_te=e\frac \kappa r\partial_r(r\partial_r\log s).
\end{equation}
No diffusion therefore occurs in the unperturbed disk, where $s$ is a
power law of the radius.

\subsubsection{Units and common parameters}
A number of parameters are common to all the runs that have been
performed. These are:
\begin{itemize}
\item The planet to star mass ratio is $q=10^{-5}$ (this translates
  into a planetary mass of $M_p=3.3$~$M_\oplus$ if the central star
  has a solar mass.)
\item The disk aspect ratio at the location of the planet is $h=0.05$.
\item As our calculations are two dimensional, we have to soften the
  planetary gravitational potential $\Phi_p$ over a length scale
  $\epsilon$:
\begin{equation}
  \label{eq:65}
  \Phi_p=-\frac{GM_p}{\sqrt{r^2+\epsilon^2}}
\end{equation}
We performed our calculations with $\epsilon=0.5H$.
This value of $\epsilon$ is meant to give results representative of
the dynamics in a slice of the disk at $|z| \sim 0.5H$. We note that
our results depend indirectly on the value of $\epsilon$, since, as we
shall see, the degree of saturation of the horseshoe region depends on
its half-width $x_s$, which is itself a function of the softening
length. A systematic study of the dependence of $x_s$ on the softening
length, as well as a study of the corresponding unsaturated torque, has
been performed by \citet{mc09}. In section
\ref{sec:transition-from-two} we add up the contributions from slices
at all $z$ to get a tentative estimate of the three dimensional torque.

\item We strictly remain in the framework either of the barotropic
  approximation, or in the framework of the study of \citet{mc09}, who
  assume a flat temperature.  In both cases, the initial temperature
  profile must be flat. In the FARGO code, the temperature profile is
  imposed through the parameter $f$ (called flaring index), defined
  by:
\begin{equation}
  \label{eq:66}
  h=h_0\left(\frac{r}{r_0}\right)^{f}.
\end{equation}
In a disk with a flat temperature profile, the parameter $f$ must be
set to $1/2$.
\item In the runs with an energy equation, the adiabatic index
  $\gamma$ of the gas is set to $1.4$.
\end{itemize}

\subsection{Sets of calculations}
\label{sec:sets-calculations}
A general situation involves a disk with arbitrary vortensity and
entropy gradients ${\cal V}_g$ and ${\cal S}_g$, and arbitrary values
of viscosity and diffusivity. In this situation, the steady state
solution for the entropy, which evolves independently of the load,
yields the amount of singular load $L_g$ that is produced at the
downstream separatrices. The general situation therefore amounts to
finding a steady state solution $L$ for the load, that fulfills
Eq.~(\ref{eq:35}) and the boundary conditions specified at
Eqs.~(\ref{eq:37}) to~(\ref{eq:39}) and~(\ref{eq:42}) with $L_0=L_g$
and $L'_\infty={\cal V}_g/a$. Owing to the linearity in $L$ of
Eq.~(\ref{eq:35}), we notice that if we have two solutions of this
equation $L_1$ and $L_2$, which respectively obey the boundary
conditions with $L'_\infty={\cal V}_g/a$ and $\Lambda_0=0$ for $L_1$
(hence ${\cal S}=0$), and $L'_\infty=0$ and $\Lambda_0=L_g$ for $L_2$,
then $L_1+L_2$ is the solution sought for the general problem. The
field $L_1$ corresponds to a barotropic situation (no load is created)
with an arbitrary vortensity gradient, while the field $L_2$
corresponds to a non-barotropic situation with a vanishing vortensity
gradient.

This allows us to split the study in two distinct parts:
\begin{itemize}
\item A study of the saturation of the vortensity related torque,
  which we perform for a globally isothermal disk. This study only
  requires varying the viscosity.
\item A study of the saturation of the entropy related torque, in a
  disk without vortensity gradient. The asymptotic torque value
  depends on two parameters: the viscosity and the thermal
  diffusivity.
\end{itemize}

The torque in a general situation can be deduced by adding the
asymptotic values of the torque in these more restricted situations.
We nevertheless comment that \citet{mc09} found a significantly larger
width (by up to $10-15$~\%) of the horseshoe region at large value of
$|{\cal S}|$, so that the entropy related torque cannot solely be put
on the account of the production of load at the downstream
separatrices. This additional complexity may render slightly
inaccurate the procedure of adding up the results of the separate
cases.

\begin{deluxetable*}{ccccccccc}
\tablecaption{Parameters used in the different series of calculations. Each series consists of $21$~runs, with $n$ ranging from $0$ to $20$. The viscosity or diffusivity therefore ranges from $2.10^{-9}a^2\Omega_p$ to $2.10^{-5}a^2\Omega_p$, in a geometric sequence. In the runs with an energy equation, whenever a parameters varies, the other is kept fixed with value $10^{-6}a^2\Omega_p$. This value was chosen to minimize the saturation due to either viscous or diffusive processes.\label{tab:sets-calculations}}
\tablehead{\colhead{Name} & \colhead{EOS} & \colhead{$\alpha$} & \colhead{$\nu/(a^2\Omega_p)$} &
\colhead{$\kappa/(a^2\Omega_p)$} & \colhead{$R_{\rm min}$} & \colhead{$R_{\rm max}$} &
\colhead{$N_{\rm rad}$} & \colhead{$N_{\rm sec}$}}
\startdata 
S$_1$ & Isothermal & $0$ & $2.10^{-9+n/5} $ & $-$ & $0.7$ & $1.3$ & $1000$ & $1500$ \\ 
S$_2$ & Energy equation & $3/2$ & $2.10^{-9+n/5} $ & $10^{-6}$ & $0.7$ & $1.4$ & $900$ & $1200$ \\ 
S$_3$ & Energy equation & $3/2$ & $10^{-6}$ & $2.10^{-9+n/5} $ & $0.7$ & $1.4$ & $900$ & $1200$ \\
\enddata
\end{deluxetable*}
 
From the above considerations, we performed three series of
hydrodynamical calculations: one in a globally isothermal disk, in
which we vary the viscosity, and two in a disk with an energy
equation, in which we respectively vary the viscosity and the
diffusivity. We sum up in Tab.~\ref{tab:sets-calculations} the
specific parameters of these series.  The mesh has a radial extent
that ranges from $R_{\rm min}$ to $R_{\rm max}$, evenly spaced in
$N_{\rm rad}$ annuli. Azimuthally, it is divided evenly in $N_{\rm
  sec}$ sectors. The calculations were ran for up to either $500$ or
$10^3$ orbits, although those which had evidently reached their
asymptotic torque value before that date were stopped manually.

We note from Tab.~\ref{tab:sets-calculations} that the isothermal
calculations are performed with a flat surface density profile (hence
${\cal V}=+3/2$). We do not have any freedom in the choice of the disk
parameters for the runs with an energy equation: since those must have
${\cal V}=0$ and since they must have a flat temperature profile, we
find that they have an entropy gradient ${\cal S}\approx 0.43$. Since
this value is positive, we expect a negative entropy related torque.

We also note that the small value of the diffusion coefficients
adopted in some of our runs raises the possibility that numerical
diffusion due to grid effects could be important. Although we have
not undertaken a systematic study of those, we are confident that
they are at most of the order of the smallest diffusivity considered here
($\sim 10^{-9}a^2\Omega_p$). Should that not be the case,
numerical diffusion would wash out the torque dependence on
viscosity or thermal diffusivity, when either of these quantities is
small. As we shall see below, we get a definite, measurable dependence
of the torque on the diffusion up the smallest value of the diffusion
coefficient (except when the flow becomes chaotic, but even though one
can perform time averages and recover a dependence on the diffusion).
This clearly indicates that the numerical diffusion present in our
runs is negligible.

\subsection{Results}

\subsubsection{Isothermal runs}
\label{sec:isothermal-runs}
\begin{figure}
  \centering
  \plotone{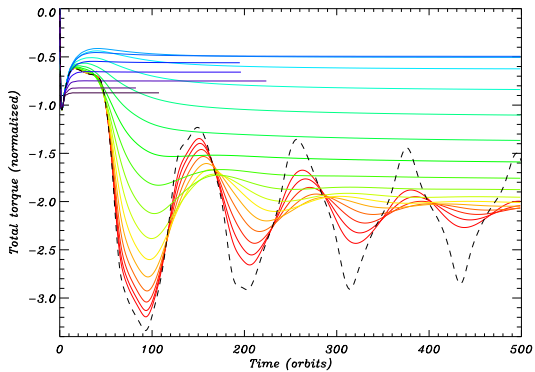}
  \caption{\label{fig:res_serie1}Total torque as a function of the
    time for the runs of set 1 (isothermal equation of state). The
    dashed line shows an additional inviscid run performed with the
    same set of parameters than the others. The viscosity ranges from
    $2\cdot 10^{-9}a^2\Omega_p$ (lower oscillatory curves, darker) to
    $2\cdot 10^{-5}a^2\Omega_p$ (upper curves, lighter). On the
    electronic version the viscosity increases as the ``wavelength''
    of the color decreases (the palette is approximately that of the
    rainbow).}
\end{figure}

The results of set~1 are displayed in
Fig.~\ref{fig:res_serie1}.  An additional inviscid run is shown, which
displays the characteristic sawtooth shape found by
\citet{2007LPI....38.2289W} in the framework of the reduced
model. Higher viscosity runs converge much before $t=500$~orbits,
while those with viscosity $\nu \lesssim 2.10^{-8}a^2\Omega_p$ still
display oscillatory behavior at that date. Evaluating the horseshoe
drag from these runs amounts to subtracting the differential Lindblad
torque from the total torque.  We have noticed that the successive
minima and maxima of the horseshoe drag, at very low but non-vanishing
viscosity, are in geometric progression, with a common ratio $r \in
]-1,0[$. Differently said, if we denote with $\gamma_1$ and $\gamma_3$
the value of two successive minima (maxima) of the total torque and
with $\gamma_2$ the value of the maximum (minimum) in between, then
the value of the differential Lindblad torque $\Delta\Gamma_{LR}$ that
we infer assuming that $(\gamma_i-\Delta\Gamma_{LR})_{i\in\{1,2,3\}}$
are in geometric progression, which is:
  \begin{equation}
    \label{eq:67}
    \Delta\Gamma_{LR}=\frac{\gamma_1\gamma_3-\gamma_2^2}{\gamma_1+\gamma_3-2\gamma_2},
  \end{equation}
  turns out to be remarkably independent of the set of successive
  extrema that we consider. This allows us to give a relatively
  accurate estimate of the differential Lindblad torque, as well as an
  estimate of the asymptotic value of the torque at low viscosity
  (which would otherwise require prohibitively long integrations).  We
  find the following value for the differential Lindblad torque:
  $\Delta\Gamma_{LR}\approx -2.14\Gamma_0$. We comment that, in order
  to evaluate $\Gamma_0$ by means of Eq.~(\ref{eq:62}), we use a value
  of $x_s$ inferred from a streamline analysis. We find $x_s^{\rm
    iso}=0.0153a$, a value that is essentially independent of ${\cal
    V}$ \citep{mc09}. The corresponding value of $\Gamma_0$ is used to
  normalize the torque value in all our isothermal runs.

\subsubsection{Runs with an energy equation}
\label{sec:runs-with-an}
\begin{figure}
  \centering
  \plotone{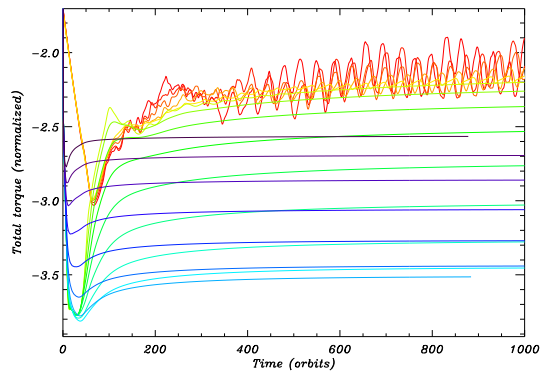}
  \caption{\label{fig:res_serie2}Total torque as a function of time
    for the runs of the set~2 (varying viscosity). The color code is
    the same as in Fig.~\ref{fig:res_serie1}: a same color refers to
    the same value of $n$. For legibility purposes, the results of
    runs $n=0$ to $n=5$ (darker curves, reddish colors on the
    electronic version) are averaged over a temporal window of width
    $120$~orbits, as they would otherwise display very large
    oscillations. For larger values of the viscosity a smooth behavior
    is observed. We note that the time averaging of the torque value
    distorts its initial behavior: without averaging, the torque of
    runs $n=0$ to $5$ would fall on top of the other curves for $t <
    50$~orbits.}
\end{figure}

\begin{figure}
  \plotone{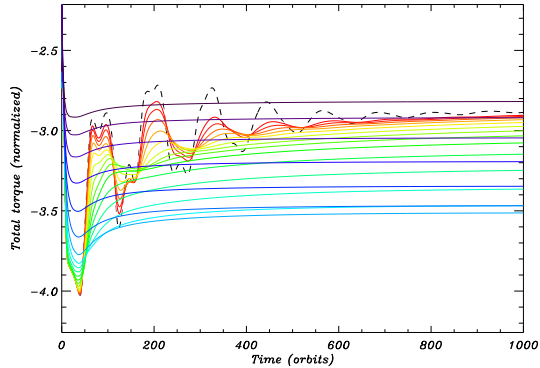}
  \caption{ \label{fig:res_serie3} Total torque as a function of time
    for the runs of the set~3 (varying thermal diffusivity). The color
    code is the same as in Figs.~\ref{fig:res_serie1}
    and~\ref{fig:res_serie2}.  No temporal averaging is performed for
    these graphs. The dashed line shows the torque in the adiabatic
    case (vanishing thermal diffusivity).}
\end{figure}

The results of set~2 and~3 are displayed respectively in
Fig.~\ref{fig:res_serie2} and~\ref{fig:res_serie3}. Calculations
performed at low viscosity (therefore in set~2, see
Fig.~\ref{fig:res_serie2}) show a highly time varying behavior which
prevents from extracting useful average values from these
calculations. We shall discuss in a little more detail this behavior
in section~\ref{sec:cons-about-invisc}.  It is straightforward to
estimate the differential Lindblad torque for these two sets: since
there is no vortensity gradient, it suffices to perform a run with the
same configuration, but with an isothermal equation of state.  The gas
being barotropic, the total torque amounts to the differential
Lindblad torque, which we divide by $\gamma=1.4$ to obtain the
differential Lindblad torque in the runs with an energy equation
\citep{bm08}. We measure $\Delta\Gamma_{LR}\approx -2.13\Gamma_0$,
where the value of $x_s$ that we adopt to estimate $\Gamma_0$ is
$x_s^{\rm iso}/\gamma^{1/4}$.

We make the following comment regarding these two sets: in an earlier
version of these calculations, we noticed a long term drift of the
torque value, which we attributed to a possible issue of boundary
conditions associated with a relative drift of the disk's material
with respect to the orbit. In order to get rid of this drift, we have
adopted, for these two sets of calculations only, a profile of the
kinematic viscosity that reads:
\begin{equation}
  \label{eq:68}
  \nu=\nu_0\left(\frac ra\right),
\end{equation}
where $\nu_0$ is the kinematic viscosity at the orbit. One can check
that this prescription yields a vanishing radial drift of the disk's
material for the surface density profile in $r^{-3/2}$ adopted
here. The runs performed with the prescription of Eq.~(\ref{eq:68})
display a better behaved torque on the long term, which allows to
measure an asymptotic torque value with a satisfactory accuracy.

We note that the curves at low thermal diffusivity in
figure~\ref{fig:res_serie3} do not pile up at the value of the
differential Lindblad torque, which is out of the range of the
vertical axis. Although this does not contradict any first principle
(since all these runs are performed with the large viscosity
$\nu=10^{-6}a^2\Omega_p$, so that they can sustain an unsaturated
corotation torque), this is somehow surprising, as one would expect to
be left with a vanishing corotation torque, when the thermal
diffusivity tends to zero, since there is no initial vortensity
gradient, and since the singular vortensity created at the downstream
separatrices tends to zero as the entropy is flattened over the
horseshoe region. This is not the case, however, and we will quantify
this effect in section~\ref{sec:vort-entr-coupl}.

\subsection{Comparison with the reduced model}
\label{sec:comp-barotr-case} 
We have performed the same set of hydrodynamical calculations
presented in section~\ref{sec:sets-calculations} in the framework of
the reduced model detailed in section~\ref{sec:simpl-model-flow}. We
show hereafter how the results of this model compare to the results of
the full hydrodynamical calculations. We have chosen $x_s$ to have the
same value as the one we measure for the runs of the full
hydrodynamical calculations, using the method of \citet{cm09}, which
is $x_s\approx 0.0153a$. The mesh of the reduced model has $N_x=4000$
and $N_y=100$, and it extends from $x=-0.3$ to $x=0.3$, as in the
hydrodynamical calculations. We add the estimate of the differential
Lindblad torque given at section~\ref{sec:isothermal-runs}, in order
to compare directly the results of full hydrodynamical calculations
and those of the reduced model.

\begin{figure}
  \centering
  \plotone{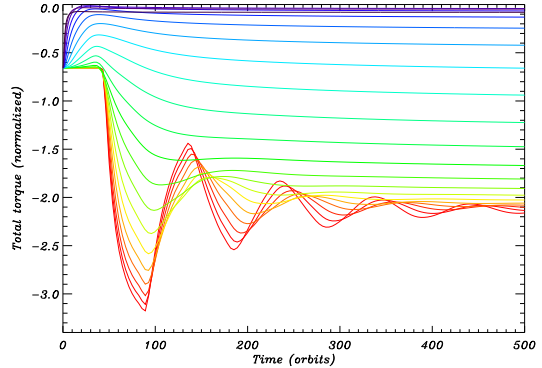}
  \caption{Torque as a function of time for the 21 calculations of the
    run set~1, performed in the framework of the reduced model.}
  \label{fig:res_toy_s1}
\end{figure}

We show in Fig.~\ref{fig:res_toy_s1} the torque as a function of time
in the barotropic case. Apart from the calculations which have a large
viscosity, these results compare very well to the full hydrodynamical
calculations (Fig.~\ref{fig:res_serie1}), which shows that most of the
long term torque behaviour observed in hydrodynamical calculations is
accounted for by the horseshoe dynamics and a simple diffusion
equation for the load. Even the slow decay of all torques with time,
still noticeable at $t=500$~orbits, is reproduced in the reduced
model. The agreement ceases when the viscosity is too large, namely
for $\nu \gtrsim 2.10^6$. For these values, the timescale of the
diffusion of load over the width of the horseshoe region, which is
$\tau_\nu\sim x_s^2/\nu$, amounts to $\sim 20$ orbits or less, that is
to say, in order of magnitude, the time required to perform a
horseshoe U-turn or to establish the horseshoe drag, as can be seen in
Fig.~\ref{fig:res_serie1}. For these large values of the viscosity,
the vortensity is therefore not conserved during a horseshoe U-turn, a
property that is not captured by the reduced model, which assumes that
the U-turns are instantaneously performed, so that they strictly
conserve vortensity, by virtue of Eqs.~(\ref{eq:38}) and
(\ref{eq:39}).

\subsubsection{Cutoff function at large dissipation}
\label{sec:cutoff-function-at}
The discrepancy at large viscosity underlined above is already
noticeable from the results over the first half libration time,
i.e. over the first $\sim 50$~orbits.  One should indeed recover the
linear estimate of the corotation torque in the limit of large
diffusion \citep{pp08}. We first entertain the barotropic case, in the
limit of large viscosity.

\begin{figure}
  \centering
  \plotone{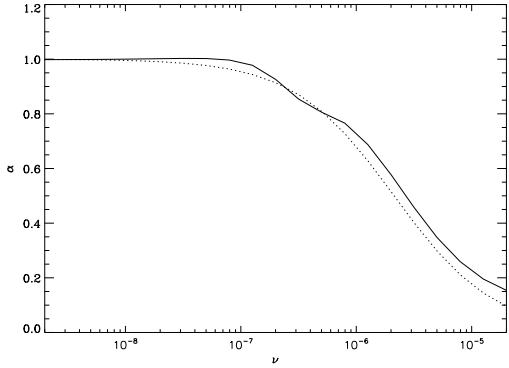}
  \caption{Blending coefficient of the isothermal series, as a
    function of viscosity. The solid curve is extracted from
    simulations using Eq.~(\ref{eq:69}), whereas the dotted line shows
    the result of Eq.~(\ref{eq:70}), with $K=1$.}
  \label{fig:cutoff1}
\end{figure}

\begin{figure}
  \centering
  \plotone{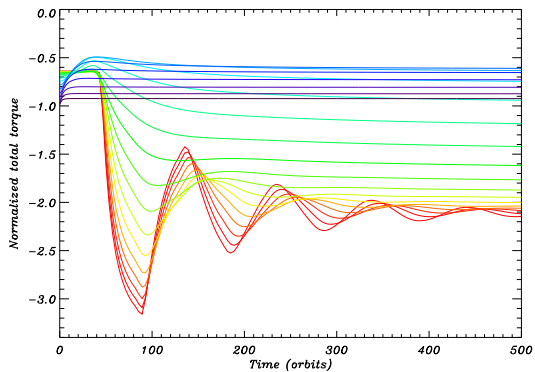}
  \caption{Total torque inferred from the reduced model, to which the
    cutoff of Eq.~(\ref{eq:70}) is applied at large viscosity. This
    figure should be compared to Fig.~\ref{fig:res_serie1}.}
  \label{fig:toyandcutoff1}
\end{figure}

In order to model in a simple manner this decay, we make the simple
assumption that for a given viscosity, the unsaturated corotation
torque is a blend of the unsaturated horseshoe drag as given by the
reduced model, and of the linear corotation torque, so that:
\begin{equation}
  \label{eq:69}
  \Gamma_{\rm CR} = \varepsilon_b\Gamma_{\rm HS}+(1-\varepsilon_b)\Gamma_{\rm CR,lin},
\end{equation}
where $\varepsilon_b\in[0,1]$ is a blending coefficient.  We can
extract from our calculations the dependency of the $\varepsilon_b$
coefficient upon the viscosity, provided we have an estimate of the
linear corotation torque. The latter can be estimated by subtracting
the differential Lindblad torque from the total torque at $t\approx
2$~orbits, which yields $\Gamma_{\rm CR,lin}\approx 1.0\Gamma_0$.  We
expect that for a sufficiently low viscosity, $\varepsilon_b \approx
1$, while $\varepsilon_b$ should tend to $0$ at large viscosity. This suggests the
following form for $\varepsilon_b$:
\begin{equation}
  \label{eq:70}
  \varepsilon_b = \frac{1}{1+K\tau_{\rm U-turn}/\tau_\nu},
\end{equation}
where $K=O(1)$ is a dimensionless coefficient, $\tau_{\rm
  U-turn}$ is the time scale for a horseshoe U-turn, and
$\tau_\nu=x_s^2/\nu$ is the viscous timescale across the horseshoe
region. We measure $\tau_{\rm U-turn}$ from the run $i=0$ of
the set~1, by evaluating the time it takes for a fluid element to go
from $x=x_s, \phi=1$~rad to $x=-x_s, \phi=1$~rad. We find
$\tau_{\rm U-turn}\approx 1.1\cdot 10^2$~$\Omega_p^{-1}$. We compare
the blending coefficient given by the simulations to the trend given
by Eq.~(\ref{eq:70}) in Fig.~\ref{fig:cutoff1}. 
We see in figure~\ref{fig:toyandcutoff1} how the correction of the
reduced model yields results which are in close agreement to those of
the full hydrodynamical calculations.
From now on, we shall consider that the results of the full
hydrodynamical simulations are correctly reproduced by the reduced
model to which we apply the cutoff provided by the blending
coefficient of Eq.~(\ref{eq:70}). This cutoff can be obtained from
short runs, that are limited to the first half libration time after
the planet is introduced in the disk.

We repeat a similar analysis for the runs with an energy equation. The
decay of the torque at large dissipation can occur either for a large
viscosity or a large thermal diffusivity. Assuming the behavior to be
separable in viscosity and thermal diffusivity, we find from the
analysis of the torque over the first $50$~orbits that the following
{\em ad hoc} functions render account of the torque cut off,
respectively at large viscosity and at large thermal diffusivity:
\begin{equation}
 \label{eq:71}
 \varepsilon_\nu=\frac{1}{1+(K\tau_{\rm U-turn}/\tau_\nu)^2},\mbox{~~~~ with
 $K=0.2$,}
\end{equation}
\begin{equation}
 \label{eq:72}
 \varepsilon_\kappa=\frac{1}{1+K\tau_{\rm U-turn}/\tau_\kappa},\mbox{~~~~ with
 $K=0.5$.}
\end{equation}

\section{Saturation functions}
\label{sec:saturation-functions}
In this section we work out suitable analytic expressions for the
perturbed vortensity and entropy fields at large time, and the
corresponding asymptotic horseshoe drag estimates. We perform our
analysis in the framework of the assumptions mentioned in
section~\ref{sec:govern-equat-reduc}, together with the boundary
conditions specified in section~\ref{sec:bound-cond-reduc}, and we
assume that the flow is steady state in the frame corotating with the
planet.

\subsection{Timescale for the asymptotic regime}
\label{sec:timesc-asympt-regime}
In a low diffusion case, it can take a very long time for the
vortensity or entropy to relax towards a steady state. For the lowest
values considered here ($\nu= 2.10^{-9}a^2\Omega_P$ or
$\kappa=2.10^{-9}a^2\Omega_p$), the time required to relax towards a
steady state over the whole mesh is:
\begin{equation}
 \label{eq:73}
 \tau_{\rm relax}\sim \frac{|R_{\rm max,min}-r_c|^2}{\min(\nu,\kappa)}\sim 10^7\mbox{~orbits}.
\end{equation}
This time, which is longer than the lifetime of the disk if $a\gtrsim
1$~AU, is much longer than the time scale accessible to fully
hydrodynamical calculations, and is still beyond the reach of reduced
models (we are able to run reduced models at most over $\sim
10^5$~orbits). In order to assess the validity of our analytic
expressions, we therefore proceed in a two step manner:
\begin{itemize}
\item We compare the results of the reduced model with the full
  hydrodynamical calculations at the final date reached by the latter,
  in order to assess the accuracy of the results of the reduced model.
\item We compare our analytic expressions with the results of the
  reduced model at much larger times.
\end{itemize}

\subsection{Saturation in the barotropic case}

We note that Eq.~(\ref{eq:34}), which describes the evolution of the
load in the barotropic case, is formally identical to
Eq.~(\ref{eq:36}), which describes the evolution of the entropy in the
baroclinic case (one has to substitute the load with the entropy, the
viscosity with the thermal diffusivity, and the gradient of load with
the gradient of entropy). This analysis therefore also provides the
information that we shall later need to evaluate the production of
load at the downstream separatrix in the baroclinic case, as it
requires the value of entropy at the separatrices.  For the sake of
definiteness, we consider the load in the remaining of this
section. Adapting the results to the entropy is straightforward.

\subsubsection{Reduction to a convolution equation at low diffusion}

If we perform the change of variable $u\equiv x^{3/2}$,
Eq.~(\ref{eq:34}) can be recast, for $x>0$, as:
\begin{equation}
 \label{eq:74}
 \frac{8A}{9}\partial_yL=\nu\frac{\partial^2L}{\partial
   u^2}+\frac{\nu}{3u}
\frac{\partial L}{\partial u}.
\end{equation} 
If we assume that $L$ varies more rapidly than $u$, Eq.~(\ref{eq:74})
can be recast as:
\begin{equation}
 \label{eq:75}
 \frac{8A}{9}\partial_yL=\nu\frac{\partial^2L}{\partial
   u^2}.
\end{equation}
Eq.~(\ref{eq:75}) is a diffusion equation in which $y$ plays the role
of the time (we note that $A<0$, and that $y$ decreases as time
evolves for a fluid particle with $x>0$). We denote with $L_\xi$ the
value of the reduced load at $y=2\pi a(1-\xi)$. We therefore have
$L_0=L^R$ and $L_1=L^F$.  For a given initial configuration $L_0$ in
$y=2\pi a$, the final configuration, after the fluid has executed a
full orbit in the corotating frame, is given by:
\begin{equation}
  \label{eq:76}
  L_1=L(y=0)=L_0\star G,
\end{equation}
where $\star$ denotes the convolution product and where the Green's
function $G$ is given by:
\begin{equation}
  \label{eq:77}
  G(u) = \frac{1}{\pi\sqrt{9\nu a/|A|}}\exp[-u^2|A|/(9\nu \pi a)]
\end{equation}
The above formulation is valid when the width of the convolution
kernel is small compared to the width of the horseshoe region
(otherwise edge effects may be important, and a significant diffusion
from the region with $x<0$, which drifts towards increasing values of
$y$, may occur). This condition translates into:
\begin{equation}
  \label{eq:78}
  \frac{9\nu a\pi}{|A|} < x_s^3
\end{equation}
We define:
\begin{equation}
  \label{eq:79}
  z_\nu = \frac{a\nu}{\Omega x_s^3},
\end{equation}
which represents, within a numerical factor, the ratio of the
libration and viscous diffusion time scales across the horseshoe
region.  The condition of Eq.~(\ref{eq:78}) therefore translates into:
\begin{equation}
  \label{eq:80}
  z_\nu< 0.026=z^c_1
\end{equation}
Similarly we define
\begin{equation}
  \label{eq:81}
  z_\kappa = \frac{a\kappa}{\Omega x_s^3},
\end{equation}
which characterizes the diffusion of entropy.  We say that a given
value of viscosity (or thermal diffusivity) falls in the regime of low
diffusion when Eq.~(\ref{eq:80}) is satisfied (or when
$z_\kappa<0.026$).  This means that a localized feature is spread over
less than the horseshoe zone width after half a libration time. When
this condition is satisfied, we can disregard the shear
and write the final distribution of the load, near the separatrix and
after a synodic period (half a libration), as a convolution product
directly in $x$, so that the Green's kernel reads:
\begin{equation}
  \label{eq:82}
  G(x) = \frac{1}{\sqrt{4\pi \nu \tau_0}}\exp(-x^2/4\nu \tau_0),
\end{equation}
where $\tau_0=4\pi a/(3\Omega_px_s)$ is half the horseshoe libration
time near the separatrix.

\subsubsection{Properties of the stationary flow at low diffusion} 
\label{sec:prop-stat-flow}
\begin{figure}
  \centering
  \plotone{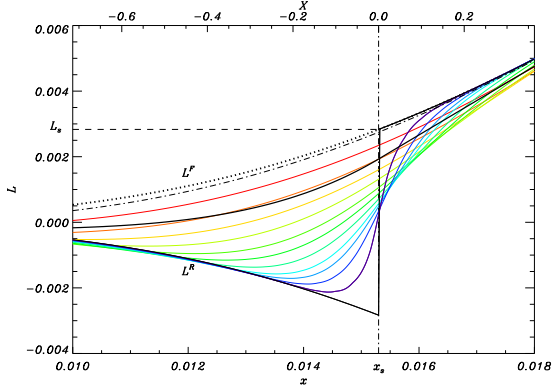}
  \caption{Perturbed load for different values of the azimuth. The
    horseshoe separatrix is at $x_s=0.0153$.  The thick solid line,
    which has a discontinuity in $x=x_s$ ($X=0$), represents the field
    just after the outward horseshoe U-turn (at the rear of the
    planet). The other curves represent the field value at different
    values of the azimuth (in geometric sequence). The thick dotted
    line shows the front load (at $y=0$, just prior to the inward
    horseshoe U-turn.) It coincides with the rear distribution (solid
    line) for $x>x_s$. The dash-dotted line shows the analytic
    expectation of Eq.~(\ref{eq:103}). The smooth solid line represents
    the azimuthal average of the load.}
  \label{fig:lowdiffsketch}
\end{figure}

\begin{figure}
  \centering
  \plotone{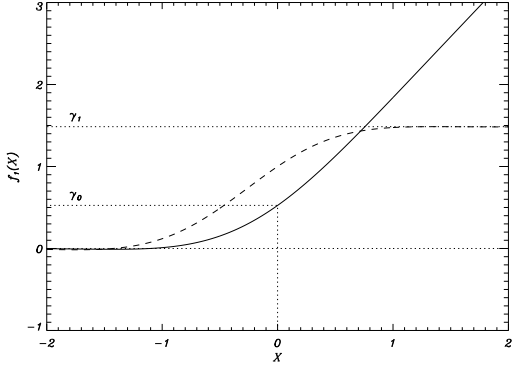}
  \caption{Normalized load distribution in the low diffusion
    limit. The solid line shows $f_1$, while the dashed line shows its
    derivative.  The horizontal dotted lines show the value of
    $f_1(0)$ and the asymptotic value of $f'(X)$ for
    $X\rightarrow\infty$.}
  \label{lowdiffnorm}
\end{figure}

Figure~\ref{fig:lowdiffsketch} shows a set of radial cuts of the load
for different values of the azimuth, in a regime of low diffusion.
One easily checks that this field obeys the boundary conditions given
by Eqs.~(\ref{eq:37}), (\ref{eq:39}), and~(\ref{eq:44}), with
$\Lambda_0=0$: the field represented by the dotted line (front field)
is the opposite of the field represented by the solid thick line (rear
field) for $x<x_s$, while they coincide for $x>x_s$. We focus our
analysis on the properties of the field in the vicinity of the
separatrix. We aim at providing a universal function that represents
the load in normalized coordinates in the low diffusion limit.  In
steady state, the field verifies the following equalities:
\begin{eqnarray}
  \label{eq:83}
  L^F &=& L^R\star G\\
  \label{eq:84}
  L^F(x)&=& \left\{ \begin{array}{ll}
      L^R(x) &\mbox{if $x>x_s$};\\
      -L^R(x) &\mbox{otherwise}.\end{array}
    \right.
\end{eqnarray}
For convenience, we define the variable \begin{equation}
  \label{eq:85}
  X=\frac{x-x_s}{\sqrt{4\nu \tau_0}}
\end{equation}
Eqs.~(\ref{eq:83}) and~(\ref{eq:84}) read, expressed in term of the
variable $X$:
\begin{equation}
  \label{eq:86}
  L^F=L^R\star K\mbox{~~~~where $K(X)=\exp(-X^2)/\sqrt{\pi}$}
\end{equation}
and
\begin{equation}
  \label{eq:87}
  L^F(X)=\mbox{sgn}(X)L^R(X).
\end{equation}
We consider the function $f_1$ that verifies Eqs.~(\ref{eq:86}) and
(\ref{eq:87}) and that obeys the following constraints:
\begin{eqnarray}
  \label{eq:88}
  f_1'(0) = 1\\
\label{eq:89}
\lim_{X\rightarrow-\infty}f_1(X)=0.
\end{eqnarray}
This last condition comes from the fact that in the low diffusion
limit, the load profile is essentially flattened over the horseshoe
region, and therefore tends to zero towards corotation (i.e. for
$X\rightarrow -\infty$).  This function can be obtained by iteration
of the transforms $T_1$ and $T_2$ respectively defined by:
\begin{equation}
  \label{eq:90}
  T_1(f):X\mapsto
\left\{
  \begin{array}{ll}
    f(X)  &\mbox{if $X>0$}\\
    -f(X)&\mbox{otherwise}
  \end {array}
\right.
\end{equation}
and
\begin{equation}
  \label{eq:91}
  T_2(f) = \frac{f\star K}{(f\star K)'(0)},
\end{equation}
where the division of the right hand side is a normalization imposed
to ensure that Eq.~(\ref{eq:88}) be satisfied. The action of $T_1$
represents the U-turns, whereas $T_2$ describes the viscous diffusion
between two subsequent U-turns. We iterate these transforms
starting from an arbitrary initial function $f$, so that iteration number $n$ is denoted $g_n$:
\begin{equation}
  \label{eq:92}
  g_n=\underbrace{T_2\circ T_1[T_2\circ T_1}_{n\;\rm times}(f)].
\end{equation}
The iterates are found to converge so that we can adopt
$f_1=\lim_{n\rightarrow\infty}g_n$. In this iterating procedure, we set $g_n$ to $0$ below an arbitrary large, negative value of $X$, in order
to satisfy Eq.~(\ref{eq:89}). 
It is straightforward to check that the couple of functions $L^F(X) = f_1(X)$ and $L^R(X) = \mbox{sgn}(X)f_1(X)$
thus constructed
satisfies
Eqs.~(\ref{eq:86})--~(\ref{eq:89}). The function $f_1(X)$ is depicted in figure~\ref{lowdiffnorm}. The actual load distribution, and ultimately the
horseshoe drag, can be obtained by a proper scaling of the function $f_1$. In order to perform this scaling, we need an additional relationship
on the radial derivative of the load.
Integrating Eq.~(\ref{eq:34}) in $y$ over the range $[0,2\pi a]$, and using the boundary condition given in
Eq.~(\ref{eq:37}), we note that:
\begin{equation}
  \label{eq:93}
  \int_0^{2\pi a}L''dy = 0\mbox{~~for any $|x|>x_s$},
\end{equation}
that is to say the azimuthally averaged slope of the field is uniform outside of the horseshoe region (even though
the timescale required to reach this state may be long, as underlined in section~\ref{sec:timesc-asympt-regime}.)
Owing to the boundary conditions given  at Eqs.~(\ref{eq:42})--(\ref{eq:43}), this azimuthally averaged slope is 
therefore necessarily:
\begin{equation}
  \label{eq:94}
  \left\langle L'\right\rangle=L_\infty',
\end{equation}
where the symbol $\langle\ldots\rangle$ denotes the azimuthal average.
We can estimate the azimuthally averaged
slope of our normalized function in $X=0$
(i.e. at the separatrix) in two different,
equivalent manners:
\begin{itemize}
\item The function $f_1(X)$ represents the load distribution in front
  of the planet ($y=0$). It does not intrinsically contain sufficient
  information to yield the azimuthally averaged slope. If we note
  $\xi=1-y/(2\pi a)$ a normalized value of the azimuth (which is $0$
  at the rear of the planet and $1$ in front of the planet), then, in
  normalized coordinates, the load distribution $f_\xi(X)$ at azimuth
  $\xi$ is given by:
\begin{equation}
  \label{eq:95}
  f_\xi = T_1(f_1) \star K_\xi=f_0 \star K_\xi,
\end{equation}
where
\begin{equation}
  \label{eq:96}
  K_\xi = \frac{1}{\sqrt{\pi\xi}}\exp(-X^2/\xi).
\end{equation}
The azimuthally averaged slope is then given by:
\begin{equation}
  \label{eq:97}
  \langle f_\xi'(0)\rangle_\xi=\int_0^1  [T_1(f_1) \star K_\xi]'(0)d\xi
\end{equation}
We note that when $\xi\rightarrow 0$, the slope can be arbitrarily
large, and that the function $f_0=T_1(f_1)$ is discontinuous in $X=0$,
so that a direct estimate of the integral of Eq.~(\ref{eq:97}) may be
inaccurate. Rather, if we note $\gamma_0=f_1(0)\approx 0.52$, then we
decompose $f_0$ as follows:
\begin{equation}
  \label{eq:98}
  f_0(X) = \mbox{sgn(X)}\gamma_0+[f_1(X)-\gamma_0]\mbox{sgn}(X).
\end{equation}
The first term of the right hand side of Eq.~(\ref{eq:98}) is a step function, whose contribution to Eq.~(\ref{eq:97}) can be evaluated
analytically, while the second term of Eq.~(\ref{eq:98}) is continuous in $X=0$, so that an accurate estimate of its contribution can
be obtained by direct summation of tabulated estimates of $[(f_1(X)-\gamma_0)\mbox{sgn}(X)]\star K_\xi$. This yields the following
estimate of the azimuthally averaged slope, that we denote with $\gamma_1$:
\begin{equation}
  \label{eq:99}
  \gamma_1= \langle f_\xi'(0)\rangle_\xi\approx 1.48
\end{equation}
\item One can notice that the slope of the load at any azimuth tends
  towards a unique value 
at large $X$ (and therefore towards the azimuthal average).
This can be inferred from examination of Fig.~\ref{fig:lowdiffsketch}, where one can see that the different curves cluster at large, positive values of $X$.
The asymptotic value of $f_1'(X)$ at large $X$ is therefore also $\gamma_1$. This is represented in Fig.~\ref{lowdiffnorm}, in which we see
directly that $\gamma_1\approx 1.48$.
\end{itemize}
The load at the upstream separatrix is scaled as follows. It reads:
\begin{equation}
  \label{eq:100}
  L^F(x) = \mu f_1\left(\frac{x-x_s}{\sqrt{4\nu \tau_0}}\right),
\end{equation}
where $\mu$ is the constant that we aim at estimating. The radial derivative of the load at the separatrix reads, using Eqs.~(\ref{eq:94}) and~(\ref{eq:99}):
\begin{equation}
 \label{eq:101}
 \left.\partial_x L^F\right|_{x_s}=\frac{1}{\gamma_1}L_\infty'
\end{equation} 
on the one hand, and, using Eqs.~(\ref{eq:85}) and (\ref{eq:88}):
\begin{equation}
 \label{eq:102}
 \left.\partial_xL^F\right|_{x_s}=\frac{\mu}{\sqrt{4\nu \tau_0}}
\end{equation}
on the other hand. Using Eqs.~(\ref{eq:101}) and~(\ref{eq:102}) we infer:
\begin{equation}
  \label{eq:103}
  L^F(x) = \frac{L_\infty'}{\gamma_1}\sqrt{4\nu \tau_0}f_1\left(\frac{x-x_s}{\sqrt{4\nu \tau_0}}\right).
\end{equation}
Figure~\ref{fig:lowdiffsketch} shows this estimates with a dot-dashed line. It compares reasonably well to the true load distribution,  which
has $z_\nu \sim 0.014$ in the example shown (half the threshold value of the low diffusion regime).
Of particular interest is the value $L_s$ at the separatrix, upstream of the U-turn.
We have:
\begin{equation}
  \label{eq:104}
  L_s= L^F(x_s) = \mu f_1(0)=L_\infty'\frac{\gamma_0}{\gamma_1}\sqrt{4\nu \tau_0}.
\end{equation}
Using Eq.~(\ref{eq:43}), Eq.~(\ref{eq:104}) can be transformed using a formal analogy to give the entropy upstream of the horseshoe U-turns, at the separatrices:
\begin{equation}
  \label{eq:105}
  S_s= S^F(x_s) = S_\infty'\frac{\gamma_0}{\gamma_1}\sqrt{4\kappa \tau_0}.
\end{equation}
This is a key quantity to evaluate the singular production of load at
the downstream separatrices, given by Eq.~(\ref{eq:48}). In the low
thermal diffusion regime, the entropy related torque therefore scales with $\kappa^{1/2}$.

\subsubsection{Case of the large diffusion regime}
\begin{figure}
  \centering
  \plotone{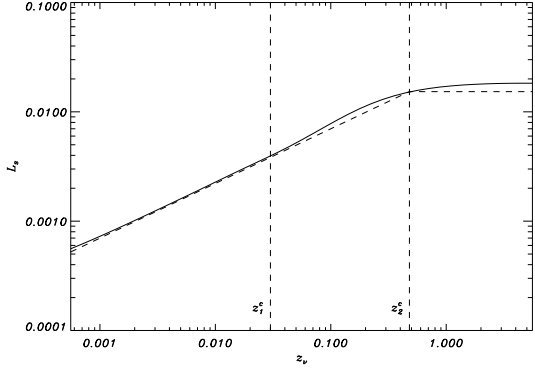}
  \caption{Load value on the separatrix, at $y=0$. The dashed line
    corresponds to Eq.~(\ref{eq:104}) up to values of viscosity that
    yield the value of the load in the unperturbed disk (i.e. for
    $z_\nu<z^c_2$), and to the value of the unperturbed disk for
    $z_\nu>z^c_2$.  The solid line corresponds to the results of
    the reduced model. We see that the analytic expression accurately
    accounts for the results of the reduced model in the low diffusion
    regime ($z_\nu<z^c_1$). For larger viscosity, some
    discrepancy is observed, with a maximal relative error of about
    $20$~\%.}
  \label{fig:loads_vs_nu}
\end{figure}

In the low diffusion regime, the value of the load at the upstream separatrix given by Eq.~(\ref{eq:104}) is much smaller than its value in the unperturbed
disk, which is 
\begin{equation}
  \label{eq:106}
L_s^0=L_\infty'x_s.
\end{equation}
This is due to the fact that the advection of load within the horseshoe region tends to flatten its profile over
the horseshoe region. Reciprocally we expect Eq.~(\ref{eq:104}) to break down when it yields values larger than the unperturbed value, which happens
for
\begin{equation}
  \label{eq:107}
  z_\nu>\frac{3\gamma_1^2}{16\pi\gamma_0^2}\approx 0.48=z^c_2
\end{equation}
Above this value we expect the diffusion to be sufficiently strong to impose at any instant in time the unperturbed load profile, so we expect
the load  at the separatrix to saturate at the value $L_s^0$. Figure~\ref{fig:loads_vs_nu} shows how this expectation compares with the
results obtained from the reduced model.
We see in this figure that the dashed line provides a reasonable approximation to the actual value of the reduced model. We shall adopt
henceforth the following approximation of the load value at the separatrix, upstream of the U-turn:
\begin{equation}
  \label{eq:108}
  L_s=L_\infty'\min\left(x_s,\frac{\gamma_0}{\gamma_1}\sqrt{4\nu \tau_0}\right)
\end{equation}
for the load in the barotropic case, and
\begin{equation}
  \label{eq:109}
  S_s=S_\infty'\min\left(x_s,\frac{\gamma_0}{\gamma_1}\sqrt{4\kappa \tau_0}\right)
\end{equation}
for the entropy value at the separatrix upstream of the U-turn, in the
case with an energy equation.

\subsection{Saturated horseshoe drag}
\label{sec:satur-hors-drag}
Integrating Eq.~(\ref{eq:34}) from $y=0$ to $y=2\pi a$ we obtain, in a steady state flow:
\begin{equation}
 \label{eq:110}
 2Ax[L^R(x)-L^F(x)]=2\pi a\nu\partial^2_{x^2}\langle L\rangle,
\end{equation}
which yields for any $|x|<x_s$, using Eq.~(\ref{eq:45}):
\begin{equation}
 \label{eq:111}
 -4Ax L^F(x)=2\pi a\nu\partial^2_{x^2}\langle L\rangle.
\end{equation}
Integrating by parts, and using the symmetry property of Eq.~(\ref{eq:44}), this allows to transform the horseshoe drag expression of Eq.~(\ref{eq:59}) into:
\begin{equation}
  \label{eq:112}
  \Gamma=8\pi B\Sigma_0\nu a^2\left[x_s\partial_x\langle
    L\rangle_s-\langle L_s\rangle\right],
\end{equation}
where we have made use of the fact that $\langle L\rangle_{x=0}=0$, which arises from the symmetry property given
by Eq.~(\ref{eq:44}). As in \citet{masset01}, we find that the
horseshoe drag reduces to a term to be evaluated at the
separatrix. The first term of the bracket of Eq.~(\ref{eq:112}) can be expressed using Eq.~(\ref{eq:94}). The second
term requires the estimate of the azimuthal average of the load at the separatrix, much in the same way as we evaluated
the azimuthal average of its derivative in section~\ref{sec:prop-stat-flow}. We obtain:
\begin{equation}
  \label{eq:113}
  \langle f_\xi(0)\rangle_\xi \approx 0.36 =
  \gamma_2\gamma_0\mbox{~~with~}\gamma_2\approx 0.68,
\end{equation}
hence we have, in the low viscosity limit, using Eq.~(\ref{eq:104}):
\begin{equation}
  \label{eq:114}
  \langle
  L_s\rangle=\frac{\gamma_2\gamma_0}{\gamma_1}L_\infty'\sqrt{4\nu \tau_0}.
\end{equation}

Using Eqs.~(\ref{eq:5}), (\ref{eq:79}), (\ref{eq:94}), (\ref{eq:112}) and~(\ref{eq:114})
we can write the horseshoe drag, in the low diffusion limit, as:
\begin{eqnarray}
  \label{eq:115}
  \Gamma &=& 8\pi B\nu a{\cal V}\Sigma_0x_s
\left[
1-\frac{4\gamma_2\gamma_0}{\gamma_1}\left(\frac{\pi}{3}\right)^{1/2}z_\nu^{1/2}
\right]\\
\label{eq:116}
&\approx&2\pi\Omega_p\nu a{\cal V}\Sigma_0x_s(1-z_\nu^{1/2}),
\end{eqnarray}
where we have restricted ourselves to the Keplerian case, and where we have used the property:
\begin{equation}
  \label{eq:117}
  \frac{4\gamma_2\gamma_0}{\gamma_1}\left(\frac{\pi}{3}\right)^{1/2}\approx 0.99\approx 1.
\end{equation}
At very low viscosity, $z_\nu\rightarrow 0$ and $\Gamma\propto\nu$. In
the low diffusion regime, Eq.~(\ref{eq:116}) can be recast as:
\begin{equation}
  \label{eq:118}
  \Gamma = \frac{8\pi}{3}{\cal V}\Gamma_0 z_\nu(1-z_\nu^{1/2}),
\end{equation}
where $\Gamma_0$ is given by Eq.~(\ref{eq:62}).
\begin{figure}
  \centering
  \plotone{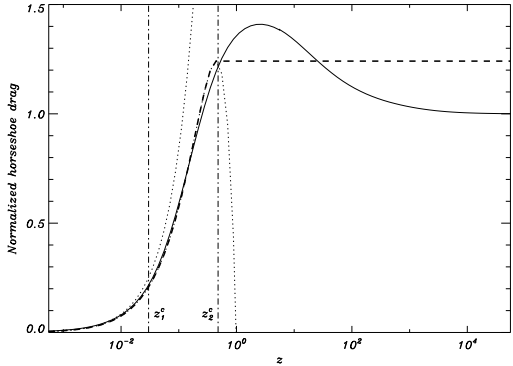}
  \caption{Horseshoe drag as a function of viscosity, normalized to
    ${\cal V}\Gamma_0$. The solid line shows the results of the
    reduced models. Note that the viscosity range has been
    significantly extended towards larger viscosities in order to
    capture the decay of the horseshoe drag past the peak around
    $z_\nu\sim 3$ and its convergence towards the standard unsaturated
    barotropic horseshoe drag given by Eq.~(\ref{eq:61}).  The upper
    dotted curve shows the linear expression $(8\pi/3){\cal
      V}\Gamma_0z$, while the lower dotted curve ---~which coincides
    with the thick dashed curve for $z<z^c_2$~--- shows
    Eq.~(\ref{eq:118}). The two vertical dot-dashed lines show the
    critical values of the viscosity that correspond to $z^c_1$ and
    $z^c_2$. The thick dashed curve shows Eq.~(\ref{eq:119})}
  \label{fig:hsdg_vs_nu_baro}
\end{figure}
Figure~\ref{fig:hsdg_vs_nu_baro} shows how Eq.~(\ref{eq:118}) compares to the actual horseshoe drag given by the calculations in the framework
of the reduced model. This expression is bound to fail at some point since it yields a negative result for $z_\nu>1$. It has a maximum
at $z_\nu=4/9$, which is $(32\pi/81){\cal V}\Gamma_0\approx 1.24\Gamma_0$. We see that the agreement between the calculations and this expression is
remarkably good up to this value of $z$ (which roughly coincides with $z^c_2$). Figure~\ref{fig:hsdg_vs_nu_baro} shows that beyond $z^c_2$, the flat approximation provides a good approximation
to the horseshoe drag, with a relative error of about $\sim 15$~\% at
most, for realistic values of the viscosity. We therefore adopt the following expression for the
barotropic horseshoe drag:
\begin{equation}
  \label{eq:119}
  \Gamma=\frac{8\pi}{3}{\cal V}\Gamma_0z_\nu F(z_\nu),
\end{equation}
where
\begin{equation}
 \label{eq:120}
 F(X)= \left\{\begin{array}{ll}1-X^{1/2}&\mbox{~~~if $X<4/9$}\\
 4/(27X)& \mbox{~~~otherwise.}
\end{array}\right.
\end{equation}
We note that, in the case of any field that obeys a dynamics
similar to that of the load in the barotropic case, i.e. that obeys
the governing equation and boundary conditions given at
sections~\ref{sec:govern-equat-reduc}
and~\ref{sec:bound-cond-reduc} (without the production of singular
stripes at the downstream separatrices),  we have the following
relationship:
\begin{equation}
  \label{eq:121}
  x_s\langle \partial_x L\rangle_s-\langle L\rangle_s=x_sL_\infty'F(z_\nu)
\end{equation}

\subsubsection{Comparison with hydrodynamical calculations and
  previous results}
\label{sec:comp-with-hydr}
\begin{figure}
  \centering
  \plotone{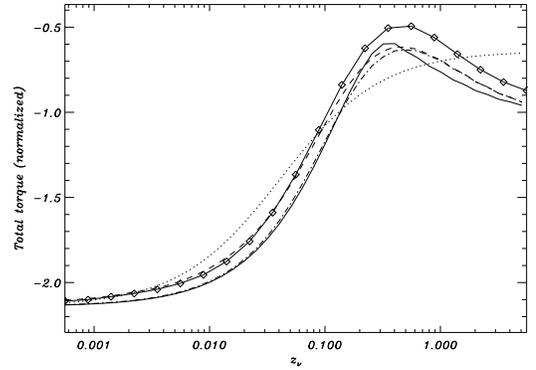}
  \caption{Total specific torque as a function of viscosity. The upper
    solid line with diamonds shows the torque at $t=500$~orbits in full
    hydrodynamical calculations.  The dashed and dash-dotted lines
    show the results of the reduced model, respectively at
    $t=500$~orbits and $t\sim 10^5$~orbits. The lower solid line shows
    the analytic expectation given by Eq.~(\ref{eq:119}), while the
    dotted line shows the analytic prediction of \citet{masset01}. The
    torque cutoff function given by Eq.~(\ref{eq:70}) is applied to
    the reduced model and to the analytic expression, except that of
    the dotted line.}
  \label{fig:tqvsnu_comp}
\end{figure}

\begin{figure}
  \centering
  \plottwo{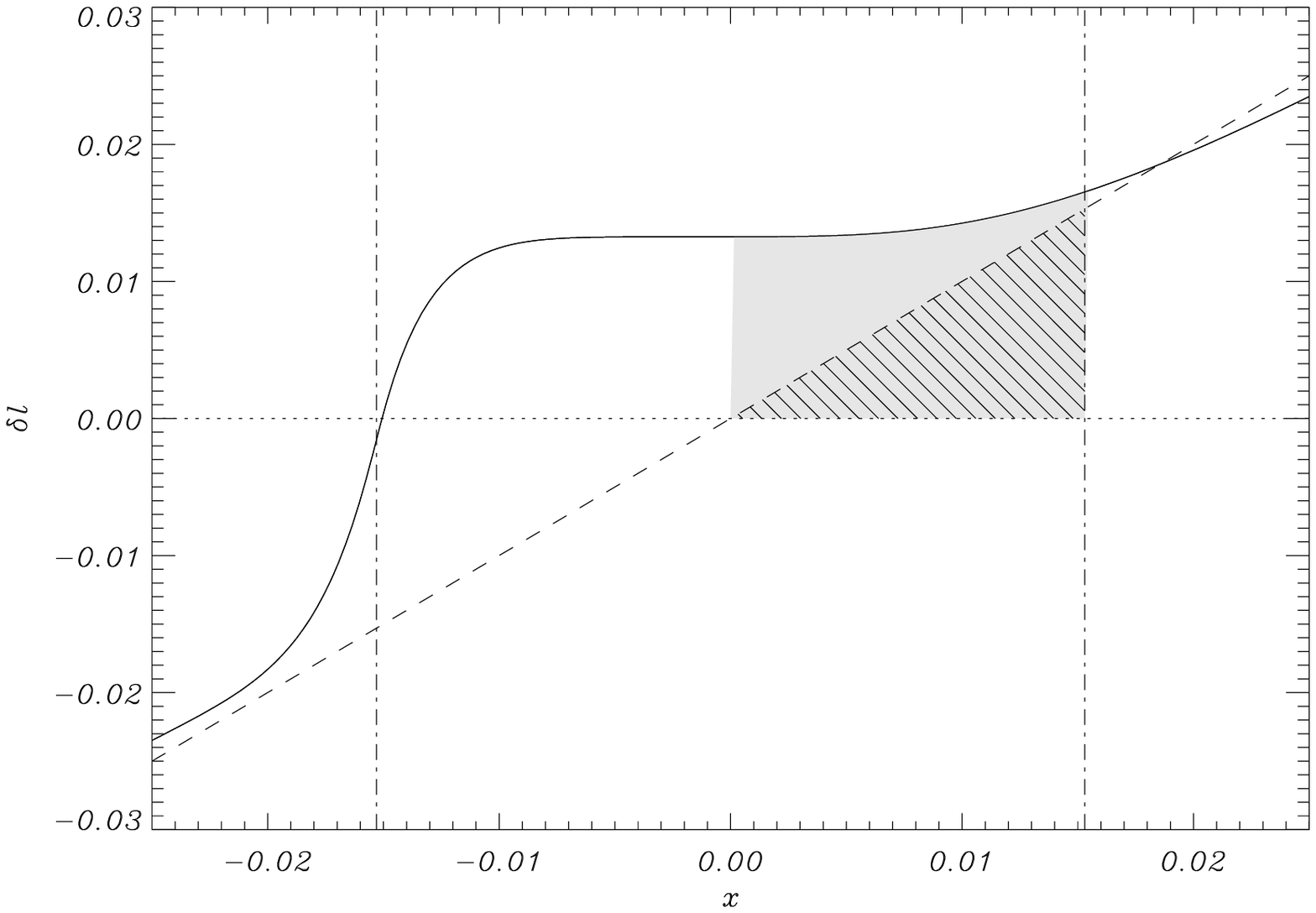}{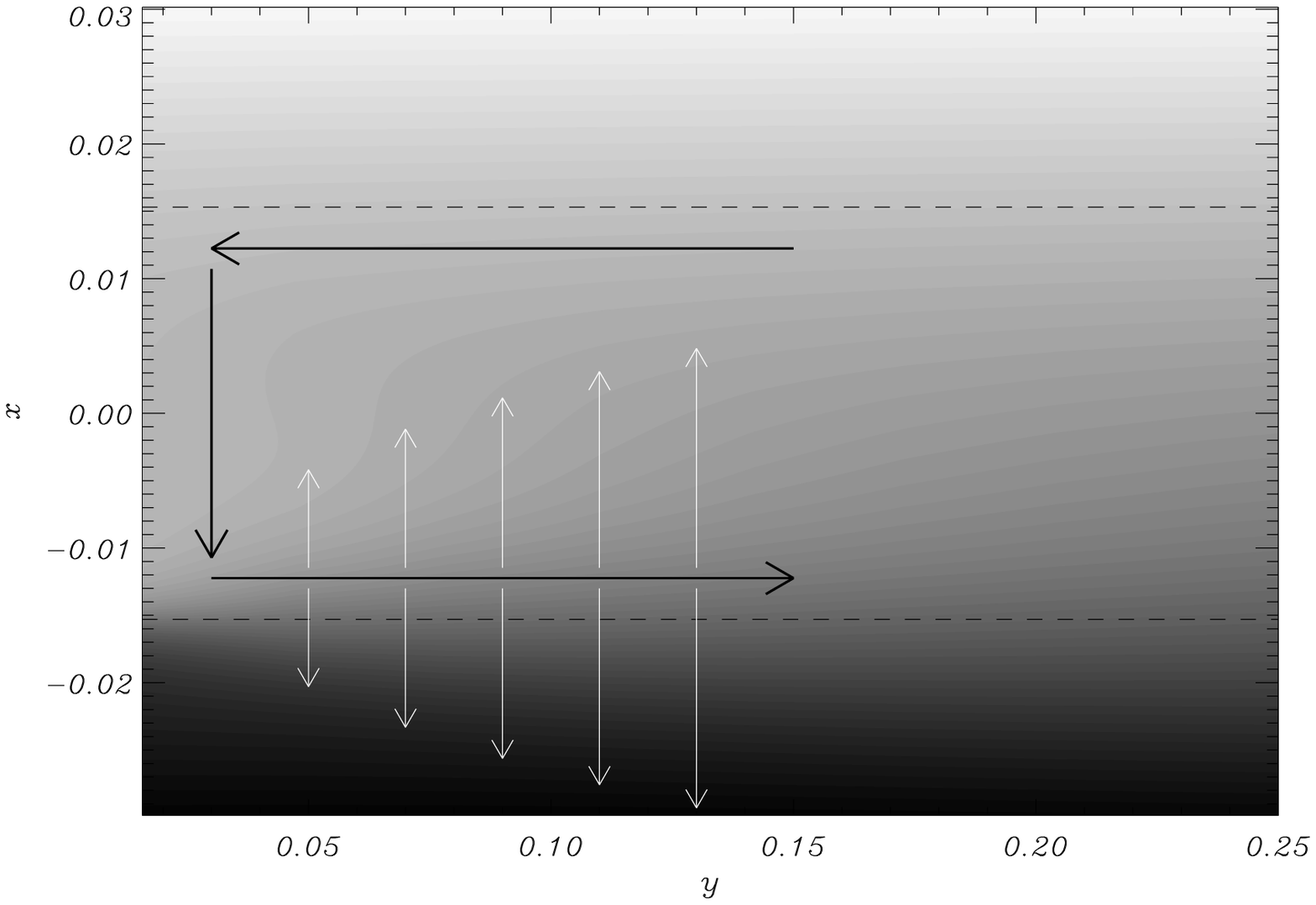}
  \caption{Load profile in front of the planet (left) and
    corresponding map of the load (right). In the left plot, the grey
    shaded area shows the actual load profile that imposes the
    horseshoe drag, while the hashed one shows the load profile that
    corresponds to the standard, unsaturated horseshoe drag given by
    Eq.~(\ref{eq:61}). In the right plot, the dark arrows sketch the
    motion of fluid, and the light arrows sketch how viscous diffusion
    spreads the load out of the horseshoe region (delimited by the
    dashed line) and to the upstream region.}
  \label{fig:supersede}
\end{figure}

We conclude this section with a comparison of the results of hydrodynamical calculations with the analytic expression, and with the expression
given by \citet{masset01}. Figure~\ref{fig:tqvsnu_comp} displays the results of full hydrodynamical calculations at a relatively early stage
($t=500$~orbits), as well as the results of the reduced model, at the same date and at a much larger time. Some difference  can be noted
between these two dates, which shows that the corotation torque at $t=500$ has not reached a steady value, in agreement with the slight decay
noticeable in the torque curves of figure~\ref{fig:res_serie1} at large time. The agreement between the full hydrodynamical calculation and the
reduced model at the same date is excellent. We note in passing that a slight ($10$~\%) rescaling has been applied to the viscosity value of the reduced model.
The saturation state of the horseshoe region depends indeed upon the ratio of the libration timescale to the viscous timescale.
This ratio is slightly larger in a real situation than in the corresponding case for the reduced model.
Indeed the horseshoe streamlines are not located at a constant distance from corotation. The existence of the potential's indirect term  renders
the streamline slightly more narrow near opposition ($\phi=\pi$). This effect is more important in thicker disks, and is marginally relevant
for the disk aspect ratio of $h=0.05$ adopted in our calculations.
This has two consequences:
\begin{itemize}
\item the libration timescale is slightly larger, as the material drifts slower near the opposition.
\item the viscous timescale is on the average slightly shorter, as the azimuthally averaged horseshoe width is smaller.
\end{itemize}
Both effects act at increasing the role of diffusion. A streamline analysis of a typical output of the hydrodynamical calculation shows that the
libration to viscous diffusion timescale is $\sim 1.1$ larger than the crude estimate of the reduced model. Hence the plots of the reduced
model results and of the analytical expectations of figure~\ref{fig:tqvsnu_comp} are performed using a value of the viscosity $1.1$ times larger
than the current viscosity, which amounts to slightly shifting leftward the corresponding graphs on the figure.
We make the following additional comments:
\begin{itemize}
\item The difference between the analytical expression and the
  hydrodynamical calculations is entirely accounted for by the date
  effect: reaching a full steady state is beyond the possibilities of
  hydrodynamical calculations, and the asymptotic torque is slightly
  more saturated than at $t=500$~orbits.
\item The expression of \citet{masset01}, even if it captures
  correctly the range of viscosities over which a transition is
  observed from full saturation to desaturation, does not quite
  provide as accurate an estimate of the torque value as
  Eq.~(\ref{eq:119}). This estimate was based upon the simplifying
  assumption that the material distribution within the horseshoe
  region has an axisymmetric distribution, which leads to an
  underestimate of the flow of load on the horseshoe U-turns. In
  addition, Eq.~(\ref{eq:119}) is much simpler to evaluate than is
  Eq.~(9) of \citet{masset01}, which involves the evaluation of Airy
  functions. Since one of the main purposes of the present work is to
  provide torque expressions that can be used in studies of planetary
  population synthesis, it is desirable to use an expression which
  minimizes the computational cost of the torque estimate.
\end{itemize}
We also notice that the steady state horseshoe drag can be larger than the unsaturated horseshoe drag, given by Eq.~(\ref{eq:61}). This is apparent
both for the reduced model and for the full hydrodynamical calculations. This behavior can be qualitatively understood as follows:
when the viscosity is large, so that the diffusion timescale across the horseshoe region is shorter than the libration timescale, the flow downstream
of the U-turns can act back on the upstream flow by diffusion, as sketched on the right plot of figure~\ref{fig:supersede}. In a steady state situation,
this can result in a load profile in front of the planet significantly above the unperturbed profile, as depicted in the left graph of figure~\ref{fig:supersede}.
This behavior also yields a value of the load at the separatrix slightly above the unperturbed one, as can be noticed in figure~\ref{fig:loads_vs_nu}.
One can estimate the excess of horseshoe drag in the situation depicted in figure~\ref{fig:supersede} by assuming that the load over the upstream
part of the horseshoe region (corresponding to the grey shaded area) is uniform and equal to the value that it has at the separatrix in the unperturbed
disk. Using Eq.~(\ref{eq:59}), it is straightforward to see that the horseshoe drag then supersedes the standard value by a factor $4/3$. This is in good agreement
with the peak value of the horseshoe drag in
figure~\ref{fig:hsdg_vs_nu_baro}. We comment that this excess is less pronounced in full hydrodynamical
calculations, since it corresponds to values of the viscosity for which some cutoff already occurs.

\subsection{Saturation of the entropy related torque}
We examine in this section the saturation properties of the entropy
related torque, and exhibit an analytic expression for the steady state entropy related
corotation torque.
Describing the saturation properties of the entropy related torque
amounts to finding the steady state pattern of load that
settles over a long time scale, when it receives a singular kick
described by Eqs.~(\ref{eq:38}) and~(\ref{eq:39}) at each horseshoe
U-turn, much as in the barotropic case.
As explained in section~\ref{sec:sets-calculations}, we split the 
general case into a bulk term that has a non-vanishing slope at 
infinity, and an edge term that has a flat profile at infinity but 
undergoes the creation of singularities at the downstream 
separatrices. We focus firstly on the bulk term, and we disregard 
the singularities.

\subsubsection{Vortensity-Entropy viscous coupling}
\label{sec:vort-entr-coupl}
For a barotropic disk we have made the simplifying assumption that the load (or
vortensity) evolution could be described by a diffusion equation, 
i.e. we have neglected the last term of Eq.~(\ref{eq:25}),
so that the evolution equation of the load was described by
Eq.~(\ref{eq:26}).
Nevertheless, the last term of Eq.~(\ref{eq:25}) may be arbitrarily
large at a contact discontinuity, and we may expect this term to play
a significant role in runs with an energy equation, at low diffusivity, when the contact discontinuity
does not significantly decay along the separatrices. Under these
circumstances, the contact discontinuity may act as a source of load,
which ultimately exerts a corotation torque. In order to work out the
contribution to the torque of the load thus created, we make two
amendments to the calculations developed in the previous section:
(i) the coupled evolution of the load and of the entropy is different,
and described by the evolution of an eigenfunction that is a linear
combination of load and entropy; (ii) the integration by parts presented in
section~\ref{sec:satur-hors-drag} must be modified accordingly, and yields a
torque expression different from that of Eq.~(\ref{eq:112}).
We briefly describe hereafter the corresponding calculations.

The evolution of the system, which is governed by Eq.~(\ref{eq:35}) and
Eq.~(\ref{eq:36}), can also be described using the eigenfunction $U$ defined by:
\begin{equation}
  \label{eq:122}
  U = L-\frac{2\nu}{\gamma(\nu-\kappa)}S.
\end{equation}
By virtue of the radial boundary conditions of
Eqs.~(\ref{eq:42}) and (\ref{eq:43}), the asymptotic slope of $U$ at
large $|x|$ is given by:
\begin{equation}
 \label{eq:123}
 \lim_{x\rightarrow\pm\infty}\partial_xU \equiv U_\infty'= \frac{\cal V}{a} -
\frac{2\nu}{\nu-\kappa}\frac{{\cal S}}{a}.
\end{equation} 
The governing equations of $U$ are therefore similar to those of $L$
in the barotropic case, except that $U$ has a different slope at infinity. Namely,
the evolution of the system is described by:
\begin{equation}
  \label{eq:124}
  \partial_tU+2Ax\partial_yU = \nu U'',
\end{equation}
together with Eq.~(\ref{eq:36}) and the boundary conditions of
Eqs.~(\ref{eq:123}) and~(\ref{eq:43}).
Also the function $U$ 
undergoes the creation of singularities $\pm L_0$ at the downstream
separatrices, as described in Eqs.~(\ref{eq:38}) and (\ref{eq:39}),
since $L$ has a unitary weight in Eq.~(\ref{eq:122}).
As explained above, we disregard these singularities in this section,
and we focus on the bulk term exclusively. The contribution of the
singular creation of load will be analyzed in the next section.

We integrate Eq.~(\ref{eq:35}) from $y=0$ to $y=2\pi a$, assuming a steady state.
This yields:
\begin{equation}
  \label{eq:125}
  -4AxL^F = 2\pi a\nu \langle L''\rangle-\frac{4\pi
    a\nu}{\gamma}\langle S''\rangle.
\end{equation}
We multiply Eq.~(\ref{eq:125}) by $8B^2ax$ , and integrate it by parts
between $0$ and $x_s$:
\begin{equation}
  \label{eq:126}
  \Gamma_{\rm HS}=8\pi Ba^2\nu\Sigma_0\left[x_s\langle
    U_s'\rangle-\langle U_s\rangle+\frac
    2\gamma\frac{\kappa}{\nu-\kappa}\left(x_s\langle
      S'_s\rangle-\langle S_s\rangle\right)\right]
\end{equation}
Using Eq.~(\ref{eq:121}), which applies to a field such as $U$, as we outlined in section \ref{sec:satur-hors-drag}, we
write:
\begin{equation}
  \label{eq:127}
  x_s\langle U_s'\rangle-\langle U_s\rangle = U'_\infty x_sF(z_\nu)
\end{equation}
and
\begin{equation}
  \label{eq:128}
  x_s\langle S'_s\rangle -\langle S_s\rangle = S'_\infty x_sF(z_\kappa).
\end{equation}
Using Eqs.~(\ref{eq:62}), (\ref{eq:123}), (\ref{eq:126}), (\ref{eq:127})
and~(\ref{eq:128}) we write:
\begin{equation}
  \label{eq:129}
  \Gamma_{\rm HS}=\frac{8\pi}{3}\Gamma_0z_\nu\left[{\cal
      V}F(z_\nu)-2{\cal S}\left(\frac{z_\nu F(z_\nu)-z_\kappa F(z_\kappa)}{z_\nu-z_\kappa}\right)\right].
\end{equation}
The last term in the bracket of Eq.~(\ref{eq:129}) represents the
additional contribution due to the vortensity-entropy viscous coupling in
Eq.~(\ref{eq:36}). By continuity, in the particular case $\nu=\kappa$ (unitary Prandtl
number), the torque is:
\begin{equation}
  \label{eq:130}
  \Gamma_{\rm HS}=\frac{8\pi}{3}\Gamma_0z_\nu\left[{\cal V}F(z_\nu)
-2{\cal S}\partial_{z_\nu}[z_\nu F(z_\nu)]\right]
\end{equation}

Eq.~(\ref{eq:130}) constitutes the bulk horseshoe drag in the general
case.
This bulk term renders account accurately of the corotation torque
that we obtain at low thermal diffusion and finite viscosity (runs set 3,
see section~\ref{sec:runs-with-an}). In this situation, the entropy
gradient virtually vanishes across the horseshoe region, so the
corotation torque essentially reduces to the bulk term, which reads,
in this particular situation with $z_\kappa/z_\nu\rightarrow 0$ and
${\cal V}=0$:
\begin{equation}
  \label{eq:131}
  \Gamma_{\rm HS}=\frac{8\pi}{3}\Gamma_0(-2{\cal S})F(z_\nu).
\end{equation}
The torque is therefore $-2{\cal S}/{\cal V}$ times that of the barotropic
situation with $\nu=10^{-6}a^2\Omega_p$, which is what one can infer
from the Figs. ~\ref{fig:res_serie1} to~\ref{fig:res_serie3}.

We now turn to the edge term which stems from the vortensity
sheets at the downstream separatrices. We comment that in order to isolate its effect
in numerical simulations, it was necessary to firstly establish
Eq.~(\ref{eq:130}) so as to subtract it from the horseshoe drag,
since, in general, Eq.~(\ref{eq:130}) shows that there is always a bulk term associated to the edge
term (i.e. when there is a non-vanishing entropy gradient), even in disks with a flat vortensity profile.

\subsection{Corotation torque associated to the singular production of
load}
As we have already considered in the previous section the bulk term in
the
general case, which stems from a non-vanishing large scale slope of
$U$, we now consider the complementary situation and isolate the effects of the singular production of load at
the downstream separatrices, and assume a vanishing large scale slope:
$U_\infty'=0$. Owing to the linearity of the flow governing equations,
the total torque is the sum of both contributions.
\subsubsection{Considerations about the inviscid case}
\label{sec:cons-about-invisc}
\begin{figure}
  \centering
  \plotone{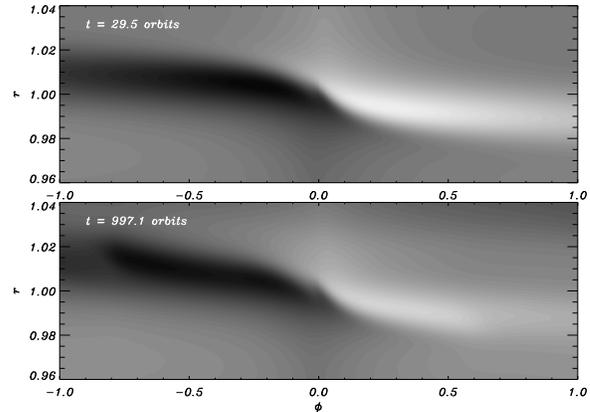}
  \caption{Perturbed entropy in a nearly inviscid disk with thermal
    diffusion. The grey level is the same for both plots. The upper
    plot corresponds to a situation during the first half libration
    time, i.e. to the fully unsaturated horseshoe drag, while the
    lower plot corresponds to approximately 12 full libration
    times.}
  \label{fig:map_entro_invisc}
\end{figure}

\begin{figure}
  \centering
  \plotone{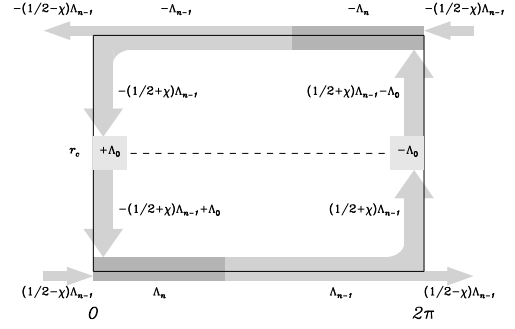}
  \caption{Sketch of the load at a given instant in time in an
    inviscid disk. The $x$-axis is the azimuth and the $y$-axis is the
  radius. The horizontal solid lines show the separatrices, while the
  dashed line shows the corotation.}
  \label{fig:sketch_invisc}
\end{figure}

The special case of an inviscid disk deserves some discussion. It has
already been noted in numerical simulations that in inviscid disks
with thermal diffusion,
the corotation torque tends to zero after several libration
timescales, so that some
viscosity is needed to prevent saturation \citep{phdbaruteau, pp08,
  2008A&A...487L...9K}. This is expected on general grounds, as
a sustained exchange of angular momentum between the horseshoe
region and the rest of the disk, described by the Navier Stokes
equations, necessarily relies on a non-vanishing stress tensor,
regardless of the thermal diffusivity
\citep{mc09,2008EAS....29..165M,2008IAUS..249..331M}.
This simple remark provides a strong clue about the origin of the
entropy related torque. It had been initially argued that this
component of the horseshoe drag came from the under- and over-dense
regions that appear at the ends of the horseshoe region as the result
of material conserving its entropy whereas the pressure remains
essentially unchanged \citep{pp08, bm08}. More recently, it has been
argued that this effect does not contribute to the entropy related
torque,
which exclusively arises from the singular production of load at the
downstream separatrices \citep{mc09}. Figure~\ref{fig:map_entro_invisc}
illustrates why this has to be the case indeed. It shows maps of perturbed
entropy corresponding to the run of set~2 with viscosity $\nu=2\cdot
10^{-9}a^2\Omega_p$ (so that it can be considered, for our purpose, as an
inviscid run). The thermal diffusivity in all the runs of this set
is such that the thermal diffusion time is longer than the horseshoe
U-turn time (so that material performing the U-turns behaves
adiabatically) but slightly shorter than half the libration timescale,
so that a fixed pattern of perturbed entropy subsists forever in the
horseshoe region (associated with a corresponding pattern of
perturbed density). Nevertheless, the corotation torque tends to zero
(on the average, since the situation is very messy owing to the
presence of vortices), as it has been seen in
figure~\ref{fig:res_serie2}. This rules out the possibility that this
localized pattern contributes, even partially, to the entropy related torque,
as was recently suggested by \citet{2009arXiv0909.4552P}. If it did,
the corotation torque could not possibly saturate at null viscosity and
finite thermal diffusion, which is in contradiction with first principles.
With this in mind,
it is instructive to undertake a simple minded analysis of the inviscid
case. We consider a planet ``switched on'' in a initially unperturbed
disk without vortensity gradient. A singular amount of load $\pm \Lambda_0$
begins to drift along the downstream separatrices. After a time $\tau_0$,
the singular load has executed a complete orbit in the corotating
frame. We can consider that the singular amount of load has spread
radially as a vanishingly narrow Gaussian under the action of a vanishingly
small viscosity. In an idealized situation, the half part of the
Gaussian packet that is outside of the horseshoe region circulates in
the outer (inner) disk, whereas the other half is shunt on a horseshoe
U-turn. In practice, however, the horseshoe region is not strictly
closed, especially at early times, and can have a slightly different
width in front of the planet and behind it \citep{cm09}. We therefore 
entertain a slightly more general situation in which a fraction
$1/2-\chi$ of the singular load circulates, whereas the remaining
fraction $1/2+\chi$ is shunt toward horseshoe U-turns, as depicted in
figure~\ref{fig:sketch_invisc}. We denote $\Lambda_n$ the load at the
front separatrix, downstream of the U-turn, over the time interval
$[n\tau_0,(n+1)\tau_0[$. The load value at this location changes
every $\tau_0$, and so does the torque by virtue of
Eq.~(\ref{eq:59}). Figure~\ref{fig:sketch_invisc} shows the
relationship between $\Lambda_{n-1}$ and $\Lambda_n$:
\begin{equation}
  \label{eq:132}
  \Lambda_n = \left(\frac 12-\chi\right)\Lambda_{n-1}
  -\left(\frac 12+\chi\right)\Lambda_{n-1}+\Lambda_0=-2\chi\Lambda_{n-1}+\Lambda_0.
\end{equation}
From this recurrence relation we infer:
\begin{equation}
  \label{eq:133}
  \Lambda_n = \left\{
    \begin{array}{ll}
      \frac{\Lambda_0}{1+2\chi}\left[1-(-2\chi)^{n+1}\right]&\mbox{if $\chi\ne -1/2$}\\
      (n+1)\Lambda_0&\mbox{if $\chi = -1/2$,}
      \end{array}
    \right.
\end{equation}
and the corresponding torque value is, using Eqs. (\ref{eq:59}) and~(\ref{eq:133}):
\begin{eqnarray}
  \label{eq:134}
  \Gamma_n =16|A|B^2ax_s^2\Lambda_0(-2\chi)^n.
\end{eqnarray}
In the general case ($|\chi|<1/2$),  the torque tends to zero at large
time. It does so after just half a libration time in the particular
case $\chi =0$. If $\chi > 0$, it oscillates about $0$, whereas it
tends monotonically to $0$ if $\chi < 0$. The ideal situation
($\chi=0$, i.e. the load is equally split at the stagnation point between
circulation and horseshoe U-turn) is more likely to be found in a
``quiet'' situation, for instance at large softening length. We expect
in that case that the torque quickly saturates (note that in practice,
the value of $\Lambda_0$ changes over time until the entropy field reaches a
steady state, so that the thermal diffusion timescale determines the
minimum amount of time needed to reach the torque saturation).

The peculiar cases $|\chi|=1/2$ are also of interest.
\begin{itemize}
\item If $\chi=1/2$ (which means that the whole amount of load
  executes horseshoe U-turns), the torque indefinitely oscillates
  about $0$. The field does not converge towards a steady
  state. However, the time averaged torque does converge to zero,
  which can be considered as a weak version of the torque saturation.
\item If $\chi=-1/2$ (which means that the load produced at the
  stagnation point never undergoes horseshoe U-turns), the load
  increases (decreases) steadily at the inner (outer separatrix). This
  rather artificial case therefore corresponds to a steady flow of
  material from one side of the horseshoe region to the other side,
  which gives rise to a non vanishing corotation torque, much as in
  type~III migration \citep{mp03}.
\end{itemize}
Therefore, apart from the artificial case $\chi=-1/2$, the torque is
bound to saturate. The manner in which it does so depends on the value
of $\chi$, hence it depends on very local properties of the flow in the
vicinity of the stagnation point. It can either oscillate
significantly before saturating, or it can tend to zero quickly and
monotonically.
This simple minded analysis therefore illustrates the fact that, even
if the asymptotic value does not depend on the details of the flow,
the manner in which the torque saturates is highly sensitive to the
flow topology, and it may differ from one particular case to the
other. This is quite in contrast with the behavior of the bulk term,
in which the torque value is dominated by phase mixing, and has a temporal
behavior much more robust and predictable.

\subsubsection{Universal form of the load in the low viscosity limit}
\label{sec:universal-form-load} 
We seek the steady state distribution of the load in the low diffusion
limit. This amounts to finding a distribution which has a vanishing
slope for $|X|\rightarrow\infty$ and which satisfies
Eq.~(\ref{eq:83}) and the following relationship:
\begin{equation}
  \label{eq:135}
  L^R (x)=-L_0\delta(x-x_s)+\left\{
 \begin{array}{ll}
      L^F(x) &\mbox{if $|x|>x_s$};\\
      -L^F(x) &\mbox{otherwise}.\end{array}
    \right.
\end{equation}
Using the variable change of Eq.~(\ref{eq:85}), we can recast these
conditions as Eq.~(\ref{eq:86}) and:
\begin{equation}
  \label{eq:136}
  L^R(X) = \mbox{sgn}(X)L^F(X)+\xi\delta(X),
\end{equation}
where $\xi=-L_0/\Delta$. We seek a couple of functions
$(L^F, L^R)$ that verifies
Eqs.~(\ref{eq:86}) and~(\ref{eq:136}) with $\xi=1$. It is
straightforward to check that if one adopts for $L^F(X)$ the function $g_1(X)$ defined by
$g_1(X)=f_1'(X)/(2\gamma_0)$,
where $f_1$ is the function that has been obtained in
section~\ref{sec:prop-stat-flow}, and for $L^R(X)$ the function given
by Eq.~(\ref{eq:136}), then the functions $L^F$ and $L^R$ so
constructed verify Eq.~(\ref{eq:86}). The function $f_1'(X)$ is
represented in Fig.~\ref{lowdiffnorm}.

For an arbitrary value of $L_0$, we have therefore:
\begin{equation}
  \label{eq:137}
  L^F(X) =\xi g_1(X) = -\frac{L_0}{\Delta}g_1(X).
\end{equation}

The horseshoe drag of Eq.~(\ref{eq:59}) can be written
\begin{equation}
  \label{eq:138}
  \Gamma_{\rm HS} = 2\int_0^{x_s}x^2L(x)dx+L_0x_s^2,
\end{equation}
which we expand, to first order in $\sqrt{4\nu\tau_0}/x_s$, as
\begin{equation}
 \label{eq:139}
 \frac{\Gamma_{\rm HS}}{8|A|B\Sigma_0a}=L_0x_s^2+2x_s^2\int_{-\infty}^0L^F(X)dX
-4\Delta L_0x_s\int_{-\infty}^0Xg_1(X)dX,
\end{equation} 
A property of $g_1(X)$ is that:
 \begin{equation}
   \label{eq:140}
   \int_{-\infty}^0g_1(X)dX=f_1(0)/(2\gamma_0)=1/2,
\end{equation}
so that the horseshoe drag expression above reduces to:
\begin{equation}
 \label{eq:141}
 \Gamma_{\rm HS}=
-32|A|B\Sigma_0a\sqrt{4\nu\tau_0} L_0x_s\int_{-\infty}^0Xg_1(X)dX.
\end{equation}
The integral on the right hand side of Eq.~(\ref{eq:141}) is
approximately equal to $-0.176$, hence we can reduce the
torque expression to, using Eq.~(\ref{eq:47}):
\begin{equation}
  \label{eq:142}
  \Gamma\approx  2.82\sqrt{\frac\pi 3} z_\nu^{1/2}\Gamma_1.
\end{equation}
The above expression, when compared to the results of the reduced
model (not reproduced here), correctly accounts for a $\nu^{1/2}$
dependence, but overestimates the horseshoe drag of the
reduced model by a significant factor (about $1.6$). The reason for
this can be tracked back to the simplification of Eq.~(\ref{eq:139})
arising from Eq.~(\ref{eq:140}), and it illustrates again the high
sensitivity of the torque edge term to the detail of the processes
that affect the load advected along the separatrix. Whereas
Eq.~(\ref{eq:140}) is exact in the framework of the set of
Eqs.~(\ref{eq:136}) and (\ref{eq:86}), which come themselves from 
Eq.~(\ref{eq:75}), it is only approximate when the shear is retained,
which means that a singular amount of load drifting along the
separatrix is not exactly split into two equal fractions on the inside and on
the outside of the separatrix. It is straightforward to estimate the
magnitude of this effect. We define the true load $L_t(X)$
distribution
as the approximate one determined above plus a residue
$\delta L(X)$ that we want to characterize:
\begin{equation}
  \label{eq:143}
  L_t(X) = L(X) + \delta L(X),
\end{equation}
and we make the assumption that $|\delta L| \ll |L|$. If we
inject this into Eq.~(\ref{eq:74}), we obtain, after integrating
from $y=0$ to $y=2\pi a$:
\begin{equation}
  \label{eq:144}
  \frac{16}{3}|A|\delta L^F = 2\pi a\frac{\nu}{u}\partial_u\langle L\rangle.
\end{equation}
The horseshoe drag of Eq.~(\ref{eq:139}) then becomes:
\begin{equation}
  \label{eq:145}
  \frac{\Gamma_{\rm HS}}{8|A|B\Sigma_0a}=2x_s^2\int_{-\infty}^0\delta
  L^F(X)dX-4\Delta L_0x_s\int_{-\infty}^0Xg_1(X)dX.
\end{equation}
In the low diffusion regime, the first integral of the right hand side
of Eq.~(\ref{eq:145}) reduces to:
\begin{equation}
  \label{eq:146}
  \int_0^{x_s}\delta L^F(X)dX=\frac{\pi a\nu}{4|A|x_s^2}\langle L\rangle_s.
\end{equation}
Using the relationship $\langle L\rangle_s=\gamma_1L^F$, we end up
with an estimate of the additional contribution to the horseshoe drag
due to $\delta L$ (the first term of the right hand side of Eq.~(\ref{eq:145}):
\begin{eqnarray}
  \label{eq:147}
  \delta\Gamma_{\rm HS }&=&16|A|B\Sigma_0ax_s^2\int_0^{x_s}\delta
  L^F(X)dX\\
  \label{eq:148}
&=&-\frac{\gamma_1}{4\gamma_0}\sqrt{\frac{\pi}{3}}z_\nu^{1/2}\Gamma_1.
\end{eqnarray}
Adding up Eqs.~(\ref{eq:142}) and (\ref{eq:148}), we eventually get
\begin{equation}
  \label{eq:149}
  \Gamma_{\rm HS}\approx 2.11\sqrt\frac\pi 3
  z_\nu^{1/2}\Gamma_1\approx 2.16 z_\nu^{1/2}\Gamma_1,
\end{equation}
which agrees with the results of the reduced model within $20$~\%.

\subsubsection{Extension to arbitrary viscosity}
Eq.~(\ref{eq:149}) remains valid as long as  the viscosity is
sufficiently small. Results obtained in the framework of the reduced
model (not reproduced here) show essentially two branches: the branch in
$\nu^{1/2}$, corresponding to the regime of low viscosity, and a flat
branch at large viscosity. These two branches intersect at $z_\nu =
0.3$, hence we adopt  the following expression for the horseshoe drag
edge term:
\begin{equation}
  \label{eq:150}
  \Gamma = \Gamma_2\varepsilon_\nu+\Gamma_{CR,lin}(1-\varepsilon_\nu),
\end{equation}
where
\begin{equation}
  \label{eq:151}
  \Gamma_2 = 2.16\sqrt{\min(z_\nu,0.3)}\Gamma_1
\end{equation}
represents the analytic trend derived in the previous section,
saturated at $z_\nu=0.3$, and where $\varepsilon_\nu$ , given by
Eq.~(\ref{eq:71}), is meant to
represent the decay towards the linear torque at large viscosity. We
note that Eq.~(\ref{eq:151}) implies, in agreement with the results of
the reduced model, an overshoot of the edge term of
about $20$~\%, much as the one we noticed in the barotropic case (see section~\ref{sec:comp-with-hydr}).

We note that so far we have assumed that the vortensity sheet produced
at the downstream separatrices has the same value as in the
unsaturated case. Actually, the horseshoe dynamics acts on the entropy
in exactly the same way as it does with the vortensity in the
barotropic case, so that the entropy jump at the stagnation point
can be lower than in the unsaturated case. It is the purpose of the
following section to check the dependence of the entropy jump (and
therefore of the torque) on the diffusivity. Here, in order to assess the correctness
of Eq.~(\ref{eq:150}) in our full hydrodynamical runs, we simply measure the entropy difference
upstream of the U-turns, in the saturated state. This difference is
the same for all runs of set~2,  and turns out to be $0.7$ times the
difference in the unperturbed disk.

\begin{figure}
  \centering
  \plotone{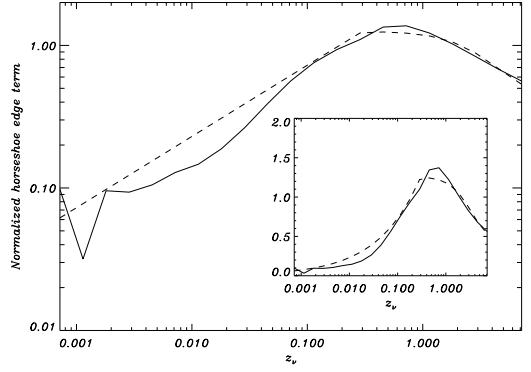}
  \caption{Entropy related torque (edge term), as a function of
    viscosity. The solid line shows the value inferred from the runs
    of set 2 (to which the estimated bulk horseshoe drag given by
    Eq.~(\ref{eq:129}) has been subtracted), while the dashed line
    shows the results of Eqs.~(\ref{eq:150})--(\ref{eq:151}). The
    value chosen for $\Gamma_{CR,lin}$ is $\Gamma_1/3$, given by
    estimates of the torque value only~$2$ orbits after the insertion of
    the planet in the disk. The inset plot represents the same
    quantity with a linear scale on the $y$-axis.}
  \label{fig:set2_vsnu}
\end{figure}

Figure~\ref{fig:set2_vsnu} shows that there is an overall correct
agreement between the results of Eq.~(\ref{eq:150})--(\ref{eq:151}), and the results of numerical simulations.
We mention at very low viscosity,  the situation is very messy owing to
the presence of vortices, as described in
section~\ref{sec:runs-with-an}, and the range of our time average (500
orbits) is too short to exhibit a neat trend. We also comment that
over the first decade of viscosities, the viscous time across the
horseshoe region is larger than the duration of the runs, so that the
corresponding runs have not necessarily reached their asymptotic value.

\subsubsection{Dependence on diffusivity}
We now examine how the horseshoe drag edge term depends on
thermal diffusion. We consider that the
dependencies of this term on viscosity and on diffusivity are separable:
its asymptotic value is the product of a function of $z_\nu$
exclusively
(which we analyzed at the previous section) by the load produced
downstream of the horseshoe U-turns (which depends on the diffusivity
exclusively, as the latter determines the entropy jump that subsists
upstream of the U-turns). We have checked this assumption with an
additional set of lower resolution runs, not reproduced here, with
diffusivity $\kappa = 3\cdot 10^{-7} a^2\Omega_p$.

We firstly note that the value of the viscosity chosen in set~3
($\nu=10^{-6}a^2\Omega_p$) corresponds nearly to the peak value in
Fig.~\ref{fig:set2_vsnu}, so that we make the simplifying assumption
that the horseshoe drag measured in these runs corresponds to the
unsaturated horseshoe drag with a lower load creation at the
downstream separatrices due to a lower entropy jump. This yields to
the torque expression:
\begin{equation}
  \label{eq:152}
  \Gamma=\Gamma_1\frac{S_s}{S_\infty'x_s},
\end{equation}
where $\Gamma_1$ is the unsaturated horseshoe drag edge term \citep[see
e.g. Eq.~(96) of][]{mc09}, and where $S_s$, the entropy value at the
separatrix upstream of the U-turn, is given by
Eq.~(\ref{eq:109}). Taking into account the cut-off at large
diffusivity, Eq.~(\ref{eq:152}) eventually yields:
\begin{equation}
  \label{eq:153}
  \Gamma\approx\Gamma_1\min\left(1,4\sqrt\frac\pi 3 \frac{\gamma_0}{\gamma_1}
z_\kappa^{1/2}\right)\varepsilon_\kappa\approx \Gamma_1\min(1,1.4z_\kappa^{1/2})\varepsilon_\kappa.
\end{equation}
We note that we do not add a blend of the linear torque value at large
diffusivity, counter to what we have done in the analysis of
dependence upon viscosity. Although we expect
to recover the linear torque value at large viscosity, we expect the
disk to behave isothermally in the limit of large diffusivity. Since
there is neither a vortensity gradient nor a temperature gradient in
the disk, we expect the corotation torque to vanish in
this limit.

\begin{figure}
  \centering
  \plotone{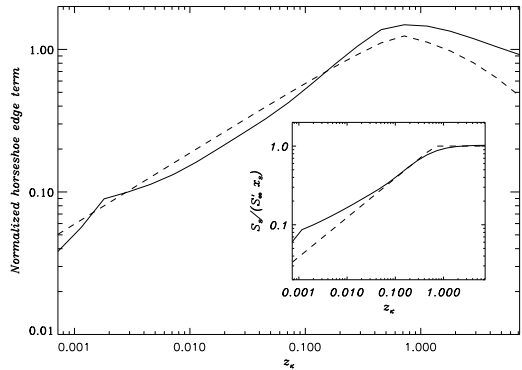}
  \caption{Entropy related torque (edge term) as a function of
    diffusivity. The solid line shows the value given by the
    simulations, corrected from the bulk term of Eq.~(\ref{eq:129}),
    while the dotted line indicates the prediction of
    Eq.~(\ref{eq:153}). The inset plot shows the ratio of the
    viscosity difference upstream of the U-turns to this difference in
  an unperturbed disk (solid line) compared to the expression of
  Eq.~(\ref{eq:109}) (dashed line). The torque values correspond to
  time averages from 800 to 1000 orbits, while the entropy jump is
  measured at $r=a\pm x_s$, $\phi=\pm 1$~rad, on a time average of the
  entropy field between $700$ and $1000$~orbits.}
  \label{fig:hsdet_vs_diff}
\end{figure}

Figure~\ref{fig:hsdet_vs_diff} shows that there is an overall correct
agreement between Eq.~(\ref{eq:153}) and the results of simulations.
We note that the entropy at very low diffusivity has not settled yet
to the expected value, as the thermal diffusion timescale is longer
than the duration of the simulations. We also mention that, whereas
the correction due to the bulk term, given by Eq.~(\ref{eq:130}), was a minute effect in the
previous section, it is a significant effect here, in particular at
low thermal diffusion where a significant viscous creation of load occurs along the
separatrices and gives rise to a relatively strong bulk term, as
already pointed out in sections~\ref{sec:runs-with-an}
and~\ref{sec:vort-entr-coupl}.

\section{Side results and additional comments}
Prior to the full torque expression, which will be given in
section~\ref{sec:gener-torq-expr}, we briefly present additional
results obtained from a systematic study of the differential Lindblad
torque, and we lay down the set of assumptions used to provide
the final torque expression provided in section~\ref{sec:gener-torq-expr}.

\subsection{Diffusivity dependence of the differential Lindblad
  torque}
\label{sec:diff-depend-diff}
For the sake of completeness, we have undertaken an additional study
of the Lindblad torque dependence on the diffusivity. At low
diffusivity, the acoustic waves that constitute the wake essentially behave as
adiabatic waves, whereas at large diffusivity, they behave
isothermally. Since the differential Lindblad torque scales as
$c_s^{-2}$, where $c_s$ is the sound speed, one should recover a
differential Lindblad torque value $\gamma$ times larger in the
isothermal limit than in the adiabatic limit. We have therefore run
calculations in which we have a vanishing vortensity gradient and a
vanishing entropy gradient (which implies $\alpha=3/2$ and $f=1/5$) so as
to have a vanishing corotation torque. These calculations are run over
a short time interval (typically 10 orbits), which is sufficient for
the Lindblad torque to reach a steady state value
\citep{MVS87}. We comment that we modified the
diffusivity procedure for these specific calculations, as we had
initially linked the amount of thermal diffusion to the Laplacian of the
entropy, in order to filter out a possible
dependence of the Lindblad torque on the diffusivity (see
section~\ref{sec:time-evol-entr}). In the present case we use the
standard dependency upon the Laplacian of the temperature. 
In the case of a single acoustic
wave in a uniform medium at rest, the transition of the phase velocity
from $c_s$ to $\gamma^{1/2}c_s$ depends on the dimensionless parameter
$\rho=c_s^2/(\omega\kappa)$, where $\omega$ is the wave frequency. This
parameter represents the ratio of the wave period to the diffusion
timescale across one wavelength. We therefore expect the transition of the Lindblad torque
to occur for a critical value of the diffusivity that is 
\begin{equation}
  \label{eq:154}
  \kappa_c  =a^2\Omega_p h ^2,
\end{equation}
obtained by letting $\rho=1$ and
adopting $\omega=\Omega_p$ for the frequency of the waves seen in the
matter
frame, since those are excited at their Lindblad resonances (where the
wave frequency matches the epicyclic frequency, hence the orbital
frequency in a Keplerian disk). We note that the characteristic
length scale in the wake excitation region is  $H$, which is also the
largest scale of turbulence, so that the diffusion approximation
should be on the edge of  its domain of validity, much as for the
horseshoe drag, as was discussed in section~\ref{sec:framework}.
This critical value of the diffusivity is much larger (by two orders of
magnitude) than the maximum diffusivity considered in our exploratory
analysis of the corotation torque. In the Lindblad torque runs, we
find that indeed the Lindblad torque has a value that is roughly the
mean of the isothermal and adiabatic limits for this value of the
diffusivity, and we provide the following fit to the dependency that we have obtained:

\begin{equation}
  \label{eq:155}
  \frac{\Gamma_{LR}}{\Gamma_{LR}^{iso}}\approx
f(\kappa/\kappa_c,\gamma),
\end{equation}
where $f(x,\gamma)$ is defined by:
\begin{equation}
  \label{eq:156}
  f(x,\gamma)=\frac{(x/2)^{1/2}+1/\gamma}{(x/2)^{1/2}+1}
\end{equation}

\subsection{Assumptions for the torque expression}
\subsubsection{Transition from two dimensions to three dimensions}
\label{sec:transition-from-two}
The study that we have presented in this paper is two dimensional,
for reasons of computational cost, so that the final torque expression
is smoothing dependent. Useful torque expressions should however be
three-dimensional, so that we scale our torque value by the
three-dimensional estimates given by \citet{mc09}. These estimates
were obtained by assuming horizontally layered motion in the coorbital
region, and by integrating the contribution of all slices, assumed to
be given, by virtue of Pythagoras'theorem, by two dimensional
estimates in which the smoothing length is the layer's altitude. This
approach, albeit not genuinely three dimensional, is more
sophisticated than a two dimensional study with a fixed, reasonable
value adopted for the smoothing length.

We
note that the saturation of the corotation torque depends on
the ratio of the libration timescale to the diffusion time scale (be
it the thermal or viscous diffusion timescale). This ratio depends on
the altitude in the disk, and its estimate requires an estimate of the
width of the horseshoe region at each altitude. Since most of the
torque arises from the layers which are located near the equatorial
plane of the disk, we simply assume that the saturation function of
the three dimensional torque can be correctly described by the
saturation of the two dimensional
torque in which we adopt the horseshoe width at low softening
length (i.e. low altitude). While this may be marginally true for the vortensity related
corotation torque and more generally for all the bulk terms, it is
likely a correct procedure for the entropy related torque and more
generally all the edge terms, the contribution of which is thought to
arise mainly from the equatorial regions \citep{mc09}.
We therefore adopt for the horseshoe width the simple value
\begin{equation}
  \label{eq:157}
  x_s=\frac 32a\sqrt{\frac qh},
\end{equation}
which is close to the maximal value found by
\citet{2009arXiv0901.2263P}. We stress that this approximate value is
required only to estimate the arguments of the saturation functions, not
the torque magnitude itself. The uncertainty on these arguments due to
the uncertainty on the value of $x_s$ is much less than the
uncertainty arising from our ignorance of diffusive processes in
protoplanetary disks.

\subsection{A note on the cut-off functions}
Accounting for the behaviour of the entropy related corotation torque at large
diffusion requires  to take into account the decay of the torque
towards its linear value \citep{2009arXiv0901.2265P}, using the cutoff
functions evaluated in section~\ref{sec:cutoff-function-at}.
We note however, as mentioned above, that we do not expect the same
value in the limit of a large viscosity (where we should recover the
linear estimate) and in the limit of a large diffusivity (where we
should get rid of the entropy related torque, as the fluid becomes
isothermal). For the sake of simplicity we make the conservative
assumption that the torque limit vanishes in both cases, motivated by
the fact that the horseshoe drag edge term should be significantly larger than the
entropy related linear corotation torque \citep{pp08}.

Regarding the cutoff of the bulk term, we use a linear vortensity related torque
value that is $2/3$ of the corresponding horseshoe drag, by comparing
the value given by \citet{mc09} in Eq.~(102) to that given by
\citet{tanaka2002}. We note that there is no need to apply a cut-off at large
thermal diffusion to the bulk term arising from the entropy-vortensity
viscous coupling, as this term is negligible in the limit
of large thermal diffusion.

\section{Generalized torque expression}
\label{sec:gener-torq-expr}
In this section we provide the torque formulae that  apply to
low mass planets embedded in viscous disks with thermal diffusion. For
the convenience of the reader mainly interested in the results rather
than in the derivation, this section is self-contained.

All the torque values given below are expressed in terms of the
canonical torque value $\Gamma_{\rm ref}$ given by:
\begin{equation}
  \label{eq:158}
  \Gamma_{\rm ref} = \Sigma_c\Omega_p^2a^4q^2h^{-2},
\end{equation}
where the reader is referred to Tab.~\ref{tab:notation} for the notation.
The total tidal torque $\Gamma_{\rm tot}$ exerted by the disk on the
planet features two components: the differential Lindblad torque $\Gamma_{\rm LR}$
and the (non-linear) corotation torque or horseshoe drag $\Gamma_{\rm
  CR}$:
\begin{equation}
  \label{eq:159}
  \Gamma_{\rm tot} = \Gamma_{\rm LR}+\Gamma_{\rm CR}.
\end{equation}
The  differential Lindblad torque is given by \citet{tanaka2002}:
\begin{equation}
  \label{eq:160}
  \frac{\Gamma_{\rm LR}}{\Gamma_{\rm ref}}=-(2.3+0.4\beta
-0.1\alpha)f\left(\frac{\kappa}{\kappa_c},\gamma\right),
\end{equation}
where $\kappa_c$ is defined by Eq.~(\ref{eq:154}) and the function $f$
is defined by Eq.~(\ref{eq:156}). 
We note that there has been some recent controversy on the
coefficient of the temperature gradient $\beta$. \citet{2009arXiv0909.4552P}
give in their Eq.~(14) a coefficient as large as $1.7$, which is
corroborated by two dimensional numerical simulations with a fixed
smoothing length of the potential. We quote here the coefficient
$0.4$ derived from the data provided by \citet{tanaka2002}. A definite answer to
this issue awaits dedicated three dimensional calculations, such as
those of \citet{gda2003}, with a systematic exploration of the
temperature gradient. Such calculations are under way, and seem
to indicate that the expression of \citet{tanaka2002} is correct, with
a high accuracy (D'Angelo, priv. comm.).

The multiplication by the function
$f$ in Eq.~(\ref{eq:160}) is intended to account for the mild
dependence of the Lindblad torque on the diffusivity. When the latter
is small enough (namely, when the relaxation timescale of temperature
disturbances is larger than the dynamical time), one may substitute
the function $f$ by the constant $1/\gamma$.

The horseshoe drag features, in general, two terms: a bulk term
$\Gamma_{\rm HS}$ and an edge term $\Gamma_{\partial\rm HS}$:
\begin{equation}
  \label{eq:161}
  \Gamma_{\rm CR} = \Gamma_{\rm HS} + \Gamma_{\partial\rm HS}.
\end{equation}
The bulk term, using Eq.~(\ref{eq:129}) and applying the normalization provided
by Eq.~(101) of \citet{mc09}, is given by:
\begin{eqnarray}
  \label{eq:162}
  \frac{\Gamma_{HS}}{\Gamma_{\rm ref}} &=& 7.8z_\nu\left[{\cal V}F(z_\nu)
-2{\cal S}\left(\frac{z_\nu F(z_\nu)-z_\kappa
    F(z_\kappa)}{z_\nu-z_\kappa}\right)\right]\varepsilon_b\\
& &+0.62{\cal V}(1-\varepsilon_b),\nonumber
\end{eqnarray}
where $z_\nu$ and $z_\kappa$ are defined respectively at
Eqs. ~(\ref{eq:79}) and~(\ref{eq:81}), and where the function $F$ is
defined by Eq.~(\ref{eq:120}). The value of $x_s$, the half width of
the horseshoe region, that is needed for the estimate of $z_\nu$ and
$z_\kappa$, is given by Eq.~(\ref{eq:157}).
Finally, Eq.~(\ref{eq:162}) also displays the bulk cutoff function
$\varepsilon_b$, given by Eq.~(\ref{eq:70}), and which can be recast,
using Eq.~(64) of \citet{bm08}, as:
\begin{equation}
\label{eq:163}
\varepsilon_b\sim (1+30hz_\nu )^{-1}.
\end{equation}
The edge horseshoe drag, using Eqs.~(\ref{eq:150}), (\ref{eq:151}),
(\ref{eq:153}) and applying the normalization provided by Eq.~(98) of
\citet{mc09}, is given by:
\begin{equation}
  \label{eq:164}
  \frac{\Gamma_{\partial\rm HS}}{\Gamma_{\rm ref}}=-3.3{\cal
    S}\,\overline{1.4z_\kappa^{1/2}}\,\varepsilon_\kappa\,\overline{1.8z_\nu^{1/2}}\,
  \varepsilon_\nu,
\end{equation}
where, using respectively Eqs.~(\ref{eq:71}) and~(\ref{eq:72}):
\begin{eqnarray}
  \label{eq:165}
 \varepsilon_\nu&\sim&[1+(6hz_\nu)^2]^{-1}\\
  \label{eq:166}
 \varepsilon_\kappa&\sim&(1+15hz_\kappa)^{-1}
\end{eqnarray}
and where
\begin{equation}
  \label{eq:167}
  \overline X \equiv \min(1,X).
\end{equation}

The set of Eqs.~(\ref{eq:159}) to~(\ref{eq:167})
provides self-contained torque formulae for a low mass planet in a
disk with arbitrary viscosity and thermal diffusivity. In the
particular case in which viscosity and thermal diffusivity remain
moderate (more precisely when $z_{\nu,\kappa}\ll h^{-1}$), one may
drop the cutoff coefficients $\varepsilon_{b,\nu,\kappa}$, which are
then close to unity.

We comment that
the estimate of the bulk term of the horseshoe
drag, given by Eq.~(\ref{eq:162}), is more accurate and robust than
the estimate of the horseshoe edge term, given by Eq.~(\ref{eq:164}),
which heavily depends on small scale properties of the flow in the
vicinity of the stagnation point of the horseshoe region, as was shown
by \citet{mc09} and as was emphasized in section~\ref{sec:cons-about-invisc}.
Also, the factor $3.3$ that features in this equation, which stems from the
conservative estimate of \citet{mc09}, is a minimum value.
In any case, this factor is large and should be sufficient to halt
migration at locations where the entropy gradient is negative and
sufficiently large in absolute value, and where the viscosity and
thermal diffusivity have adequate values to prevent saturation. As a
consequence, a minor change in the value of this coefficient should have
little qualitative impact on scenarios of migration.

\section{Discussion}
\label{sec:discussion}
\subsection{On the planetary mass range relevant to this analysis}
The torque expressions provided here are strictly valid in the
regime of small planetary mass, for which the torque scales as the
square of the planet mass (and the horseshoe width scales as the
square root of the planetary mass). \citet{mak2006} have shown, by
means of a two dimensional analysis, that when the planetary Hill radius represents
some fraction of the disk pressure length scale (typically such that
$q/h^3\sim 0.6$ in their study), the horseshoe zone width is much
larger than predicted by the $q^{1/2}$ scaling, extrapolated from
lower masses, thereby boosting the corotation
torque. \citet{2009arXiv0901.2263P} have shown that this behavior is
due to the wake, which, owing to its proximity, affects the enthalpy
at the horseshoe stagnation point, hence it affects the horseshoe
width. They found this behavior to be smoothing dependent: the
mass at which the horseshoe drag boost is observed depends on the
softening length, and is smaller at smaller softening length. If one
translates this in terms of altitude in a three dimensional disk,
this suggests that the horseshoe region is always over wide in the
vicinity of the
equator,   over a vertical length scale that increases when the
planetary mass does. The vertically integrated bulk horseshoe
drag should then be boosted when the vertical extent of the over wide
region represents a significant fraction of the disk thickness. This
expectation yields a relationship similar to that quoted above,
i.e. one expects the boost to occur for $q/h^3 \sim O(1)$, where the
dimensionless number of the right hand side is still to be
determined by three dimensional calculations. We comment however
that the edge term of the horseshoe drag, which is biased towards
low altitude regions owing to its dependence in $\epsilon^{-1}$
\citep[see][]{mc09} should be boosted at planetary masses even lower
than the bulk term. This constitutes another reason why the edge horseshoe
drag given by Eq.~(\ref{eq:164}) should be regarded as a
conservative estimate, even for planetary masses significantly smaller than
$0.6h^3M_*$.

\subsection{Relevance of a diffusion equation to model the entropy
  evolution}
\label{sec:relev-diff-model}
We have modelled the radiative processes at work in the disk by a
simple radial diffusion equation on the entropy. The merits of this
simple approach have been discussed in
section~\ref{sec:time-evol-entr}. The actual processes that contribute
to the energy balance of fluid elements in the disk, however, cannot
be reduced to this simple minded approach. \citet{2008A&A...487L...9K}
have for instance performed numerical simulations that take into
account, in addition to the processes that can be described as a
diffusion, the local processes that affect the energy balance,
i.e. the viscous heating and the radiative losses through the disk
photospheres. If one neglects the diffusion and takes into account
only local processes, then an alternate simple model of thermal
processes can be achieved by imposing a local temperature prescription,
together with a characteristic decay time of the temperature
disturbances \citep[e.g.][]{phdbaruteau}. What matters in assessing
the magnitude of the entropy-related horseshoe drag is the value of
the entropy at the separatrices upstream of the U-turns. The latter
results, in steady state, from a balance between libration and
relaxation. We note that the dimensionless variable $z_\kappa$, which
is used to quantify the corotation torque saturation due to thermal
processes, scales with the ratio of the libration time to the
thermal timescale $\tau_{\rm th}=x_s^2/\kappa$:
\begin{equation}
  \label{eq:168}
  z_\kappa = \frac{3}{4\pi}\frac{\tau_0}{\tau_{\rm th}}.
\end{equation}
Although we have not undertaken a systematic exploration of the torque
saturation in the framework of a temperature prescription such as the
one of \citet{phdbaruteau}, we note that his results are compatible
with the formulae of Eq.~(\ref{eq:164}) in which one
adopts the value of $z_\kappa$ given by Eq.~(\ref{eq:168}). In a more
general manner, we suggest that Eq.~(\ref{eq:168}) be systematically
used to assess the degree of saturation of the entropy-related
corotation torque, and that $\tau_{\rm th}$, the relaxation timescale
of temperature disturbances of scale length $x_s$, be determined by a prior
analysis of the relevant thermal processes.

\subsection{Planet-disk relative drift}
\label{sec:planet-disk-relative}
In the calculations performed for the analysis of the entropy related
horseshoe drag, we have imposed a power law for the kinematic
viscosity that yields no radial drift in the unperturbed disk. We do
not expect that a relative drift of the disk and planet, given either
by the large scale accretion of the disk's material onto the primary,
or migration, or an admixture of both processes, would alter our
conclusions, for the following reasons:
\begin{itemize}
\item The planet-disk drift turns out to be unimportant in the
  barotropic case. By extension, it should be of little importance for
  the bulk term of the horseshoe drag in the general case. 
\item The viscous drift of the vortensity sheets at the separatrices,
  over a synodic period, typically amounts to $\tau_0\nu/a$, whereas
  its viscous spread reads $\sqrt{4\nu\tau_0}$. The ratio of these two
  quantities is $\nu\Omega_p^{-1}/ax_s$, which should be small in any
  realistic disk and for any low mass planet for which planetary
  migration is relevant.
\item We can also work out the typical  amount of drift due to type~I
  migration, at a typical speed $\dot a$ given by
  \begin{equation}
    \label{eq:169}
    \dot a\sim \frac{2\Gamma_{\rm ref}}{\Omega_p M_pa}
  \end{equation}
  The drift induced by migration over a synodic
  period $\tau_0\dot a$ is therefore:
  \begin{equation}
    \label{eq:170}
    \tau_0\dot a\sim \frac 83 \frac{\mu_d}{h} x_s,
  \end{equation}
  where $\mu_d=\pi\Sigma_0a^2/M_*$ is the reduced disk
  mass. Eq.~(\ref{eq:170}) shows that the drift is a small fraction of
  $x_s$ except in very massive disks, close to their gravitational
  stability limit. Since the drift is a tiny fraction of the horseshoe
  zone width, it may impact the degree of saturation only in the cases
  for which the viscous spread of the vortensity sheets is also a tiny
  fraction of the horseshoe width, that is to say when the entropy
  related torque is significantly saturated.
\end{itemize}

\section{Summary}
\label{sec:summary}
We have undertaken a study of the corotation torque saturation, as a
function of viscosity and thermal diffusion, and we have obtained
asymptotic torque expressions that are given in
section~\ref{sec:gener-torq-expr}. A property that is essential to our
derivation is that the horseshoe drag, whether barotropic or not,
depends exclusively on the distribution of vortensity within the
horseshoe region, so that our analysis essentially reduces to finding
the vortensity distribution for a steady flow in the corotating
frame. In particular, the contribution of the under- or over-dense
regions that appear within the horseshoe region should not be
added ``manually'' to the horseshoe drag estimate.
Doing so leads to a violation of first
principles, as is discussed in section~\ref{sec:cons-about-invisc}.
Our analysis shows that the saturated horseshoe drag can be
decomposed into a bulk term and an edge term. The latter scales with
the entropy gradient, as in an unsaturated situation, and its
saturation properties depend both on the viscosity and on the thermal
diffusivity. The bulk term, however, not only depends on the vortensity
gradient, as in the unsaturated case, but also on the entropy
gradient, as a result of the viscous production of vortensity at steep
entropy gradients.

We have also contemplated in 
section~\ref{sec:diff-depend-diff} the transition from the isothermal
to adiabatic differential Lindblad torque (the latter is a factor
$\gamma$ than the former), and found that it occurs when the thermal
timescale is of the order of the dynamical timescale.
The dependence of the entropy
related torque on the thermal timescale is regulated by a markedly
different ratio, namely the ratio of the thermal timescale to the
libration timescale, which gives the degree of saturation. 

At large viscosity or thermal diffusivity, the horseshoe drag is found
to decay. We have made an attempt to account for this decay by the use
of {\em ad hoc} simple analytical fits (cutoff functions). A detailed study of the
intermediate regime that settles at large diffusion (be it viscous or
thermal), and that yields a torque value comprised between the
horseshoe drag and the linear corotation torque could be useful for
sub-terrestrial mass objects embedded in disks with significant
viscosity or short radiative timescales. One can estimate the value
required for the viscosity or thermal diffusion to recover the linear
regime. Writing that $\varepsilon_{b, \nu,\kappa}\ll 1$, we are left
with, using Eqs.~(\ref{eq:163}), (\ref{eq:165}) and (\ref{eq:166}):
\begin{equation}
  \label{eq:171}
  \frac{\nu\mbox{ or }\kappa}{a^2\Omega} \gg 0.1q^{3/2}h^{-5/2}.
\end{equation}
For the condition on viscosity, this can be translated into a lower limit
on the $\alpha$ coefficient of the disk:
\begin{equation}
  \label{eq:172}
  \alpha \gg 0.1q^{3/2}h^{-9/2}.
\end{equation}
Eqs.~(\ref{eq:171}) and (\ref{eq:172}) show that the corotation torque
should approach its linear value for an Earth mass planet in a disk
with $h=0.04$, invaded by magnetorotational turbulence (the critical
$\alpha$ value of the right hand side of Eq.~(\ref{eq:172}) being then
$10^{-3}$). If the planet has a higher mass, if the disk is
thinner, or if the planet is in the dead zone, the corotation torque
should be determined by a horseshoe drag analysis.

Finally, we have underlined that the
horseshoe drag edge term is potentially extremely strong in a three
dimensional case, so that the estimate that we have given in
Eq.~(\ref{eq:164}) should be regarded as conservative. This stresses
the need for a three dimensional study, that would allow a calibration
of the unsaturated edge term value.

\acknowledgments

The numerical simulations performed in this work have been run on a
140 core cluster funded by the program {\em Origine des Plan\`etes et
 de la Vie} of the French {\em Institut National des Sciences de
 l'Univers}. Partial support from the COAST project ({\em
 COmputational ASTrophysics}) of the CEA is also acknowledged. F.M.
also wishes to thank G. Koenigsberger for hospitality at the
{\em Instituto de Ciencias F\'\i sicas} of UNAM, Mexico.

\end{document}